\documentclass{aa}
\usepackage{graphicx}
\usepackage{natbib}
\usepackage{txfonts}

\usepackage[figuresright]{rotating}
\usepackage{supertabular}
\bibpunct{(}{)}{;}{a}{}{,}%

\font\math = cmmi10
\def\m#1{\hbox{\math \char'#1}} 
\def\v{\m{166}}

\font\smallmath = cmmi6
\def\sm#1{\hbox{\smallmath \char'#1}} 
\def\sv{\sm{166}}

\def\H2O{H$_2$O}

\begin{document}
   \title{Observational evidence for the shrinking of bright maser spots}

   \author{Anita Richards\inst{1}
          \and Moshe Elitzur\inst{2}
          \and Jeremy Yates\inst{3}
          }

   \offprints{}

   \institute{Jodrell Bank Centre for Astrophysics, School of Physics and Astronomy, Alan Turing Building, University of Manchester, M13 9PL, UK \\
              \email{\tt amsr@jb.man.ac.uk}
         \and University of Kentucky, Department of Physics and Astronomy, 600 Rose Street, Lexinton, KY, 40506-0055 USA\\ \email{\tt moshe@pa.uky.edu}
\and Department of Physics and Astronomy, University College London, Gower 
Street, London. WC1E 6BT, UK \\ \email{\tt jyates@star.ucl.ac.uk}
             }

   \date{Received September 15, 3000; accepted March 16, 3000}


  \abstract
 {The nature of maser emission means that the apparent angular size of an individual maser spot is determined by the local amplification process as well as by the instrinsic size of the emitting cloud. Highly sensitive radio interferometry images made using MERLIN spatially and spectrally resolve water maser clouds around evolved stars.}
 {We used measurements of the cloud properties, around the red supergiant S Per and the AGB stars IK Tau, RT Vir, U Her and U Ori, to test maser beaming theory. In particular, spherical clouds are expected to produce an inverse relationship between maser intensity and apparent size, which would not be seen from very elongated (cylindrical or slab-like) regions.}
 {We analysed the measured properties of the maser emission in order to estimate the saturation state. We analysed the variation of observed maser spot size with intensity and across the spectral line profiles.}
 {Circumstellar masers emanate from discrete clouds from about one to 20 AU in diameter depending on the star. Most of the maser features have negative excitation temperatures close to zero and modest optical depths, showing that they are mainly unsaturated. Around S Per and (at most epochs) RT Vir and IK Tau, the maser component size shrinks with increasing intensity, although in some cases the slope is shallower than predicted, probably due to shape irregularities and the presence of velocity gradients within clouds. In contrast, the masers around U Ori and U Her tend to increase in size, with a larger scatter.}
 {The water masers from S Per, RT Vir and IK Tau are mainly beamed into spots with an observed angular size much smaller than the emitting clouds. The brighter spots at the line peaks are smaller than those in the wings. This suggests that the masers are amplification-bounded, emanating from clouds which can be approximated as spheres. Many of the masers around U Her and U Ori have apparent sizes which are more similar to the emitting clouds and have less or no dependence on intensity, which suggests that these masers are matter-bounded. This is consistent with an origin in flattened clouds and these two stars have shown other behaviour indicating the presence of shocks which could produce this effect. }


   \maketitle
%

\section{Introduction}

Maser regions are comprised of many bright, compact spots, each with its own
well-defined Doppler velocity. Geometry is an important factor in controlling
the brightness of these individual features. \cite*{Elitzur92a}, hereafter
EHM92, suggest that the main difference between the high brightness
temperatures of \H2O masers in star-forming regions and their lower brightness
counterparts in evolved stars is the geometry---the pumping scheme and physical
conditions are similar in both but the former are characterised by planar shock
geometry while the latter are three-dimensional structures that can be modelled
as spheres. The spherical maser radius is always limited because loss lines of
the pump cycle become optically thick for sufficiently large dimensions, and
the maser thermalizes. In contrast, there are no similar limitations on the
size of either disk or filamentary masers.  In these cases the pumping cycle
thermal photons can escape along a short dimension perpendicular to the line of
sight while the length along the latter can increase arbitrarily without any
effect on the basic inversion efficiency, allowing extremely high brightness
temperatures.

\begin{figure}
   \centering
   \includegraphics[angle=0, width=9cm]{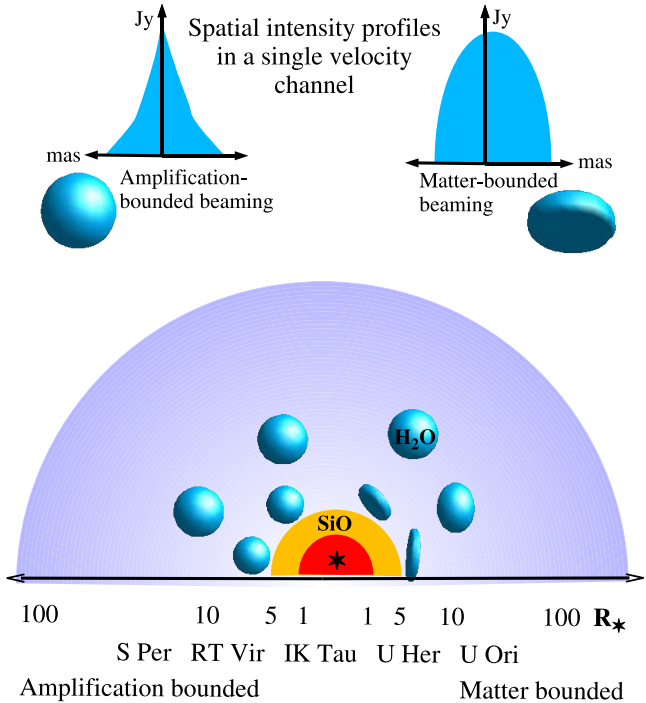}
      \caption{The lower part of this cartoon shows, on the left,
      spherical water vapour clouds in a CSE. On the right some clouds are
      flattened.   The upper diagrams show the spatial
      profiles  resulting from maser propagation through a sphere
      (\emph{left}) and through a
     disc seen edge-on (\emph{right})
The scale in terms of stellar radii is notional.   The similarities
      with the named sources are explained later in this paper.}
         \label{beaming.ps}
   \end{figure}

Unfortunately, maser geometry cannot be determined directly. As noted
in EHM92, from any given direction a spherical maser appears identical
to a cylindrical maser aligned along the line of sight. While the
length of this equivalent cylinder is the sphere's diameter, its cross
section, the spherical maser observed area, is controlled by the maser
beaming and determined by the amplification optical depth
(``amplification-bounded" maser). The observed area of an
amplification-bounded maser varies inversely with the amplification
along the line of sight---the stronger is the amplification, the
smaller is the observed area. Because the amplification decreases with
frequency shift away from line centre, the appearance of a spherical
maser varies across the line profile: the observed area is smallest at
line centre, increasing toward the line wings \citep{Elitzur90}. In
contrast, the observed area of a cylindrical maser is its actual cross
section (``matter-bounded'' maser), independent of frequency.
Therefore, the maser geometry can be determined indirectly by
examining its observed size at different frequencies across the line
profile.  The cartoon Fig.~\ref{beaming.ps} illustrates these
properties. The spherical clouds on the lower left give rise to the
amplification-bounded spatial profile shown at the upper left.  Some
of the clouds shown at lower right are flattened and, if observed
edge-on, matter-bounded masing occurs, giving rise to the spectral
profile at upper right.  

In accordance with the EHM92 proposal, \cite{Genzel81} report that
observed sizes of individual \H2O maser spots in W51M are roughly
constant across the line profiles, and similar results were obtained
for a number of other sources by R.\ C.\ Walker \& J.\ M.\ Moran
(private communication); the frequency independence of the observed
sizes of bright masers in star-forming regions conforms with the
matter-bounded geometry expected behind shock fronts.

On the other hand, stellar wind masers are expected to be local
regions carved out of the outflow by the requirements of velocity
coherence along the line of sight. The resulting geometry is in
general fully three-dimensional and amplification--bounded, thus the
observed area should increase away from the line centre. Here we test
this hypothesis for \H2O masers in circumstellar envelopes (CSE) of
evolved stars.

\H2O masers at 22 GHz are found around post-main-sequence oxygen-rich
asymptotic giant branch (AGB) and red supergiant (RSG) stars, with surface
temperatures of 2500--3500~K. AGB stars supporting water masers have radii of
$\sim 0.5-3$ AU and undergo mass loss at $10^{-6}-10^{-5}$ M$_{\odot}$
yr$^{-1}$; RSG are around an order of magnitude larger and more prolific.
The masers  occur within five to a hundred stellar radii ($R_{\star}$) of the
star. Individual features typically span 1--2 km s$^{-1}$ and the observed size
in a single 0.1 km s$^{-1}$ channel is typically a few milli-arcsec (mas).

 \begin{figure}
   \centering
   \includegraphics[angle=-90, width=9cm]{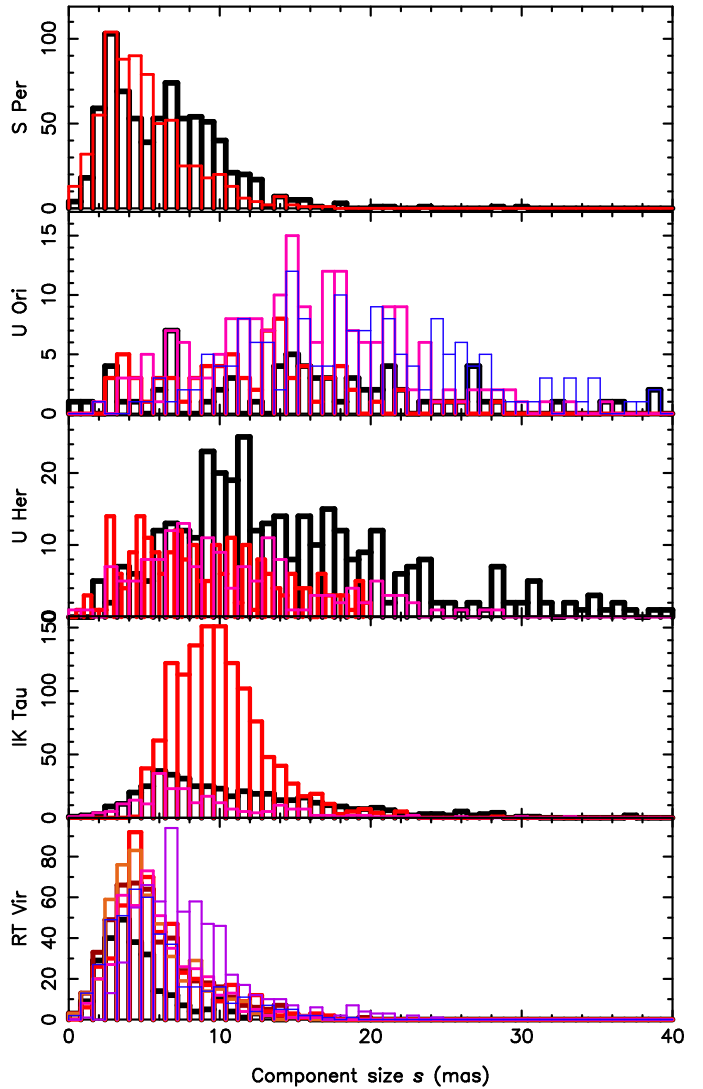}

\caption{Histogram of the sizes of H$_{2}$O maser components exceeding the
observational error ($s_{\sv}>\sigma_{s}$). The ordinate shows the
number of components per size bin.  The thickest black line shows the
distribution for the epoch with the brightest maximum emission
($I_{\mathrm{max}}$, given in Table~\ref{obstab}) for each star, with
thinner and paler/bluer lines representing epochs with progressively fainter
maxima. A few U Her and U Ori components with sizes $> 40$ mas are not
shown.   }
         \label{SPOTHIST.PS}
   \end{figure}

MERLIN\footnote{The UK radio interferometer, operated by the
University of Manchester on behalf of STFC} can produce detailed
images of H$_{2}$O maser shells around AGB stars within a few hundred
pc, or RSG within a few kpc from Earth, without losing significant
emission due to missing interferometer spacings.  We measured the
sizes of individual maser spots by fitting Gaussian components,
allowing us to investigate the variations of maser size with
intensity, position in the line profiles and other factors. The sample
and observational results are described in Section~\ref{obs}. We
summarise the evidence for maser clumping into discrete clouds and
their properties and saturation state in
Section~\ref{cloudproperties}.  The variations of the observed size of
maser components with intensity and with position in the spectral line
profile of clouds are analysed in Section~\ref{beaming}.
This provides
diagnostics for the two different maser geometries, explained 
in Section~\ref{discussion}.  Section~\ref{conclusions} summarises
compelling evidence that the variations of observed maser sizes across
 line profiles can be used to distinguish between matter-bounded and
 amplification-bounded masers.


\section{Observational measurements}
\label{obs}

\subsection{Observations}
\label{sub:obs}
We have carried out a program of MERLIN observations of 22-GHz H$_2$O
masers around evolved stars, from 1994 onwards. Four AGB stars (U
Her, U Ori, IK Tau and RT Vir) and one RSG (S Per)
have good enough visibility plane coverage to
resolve the individual maser components. Their relevant properties are
summarised in Table ~\ref{stars}.

\begin{table}
\begin{tabular}{lclccc}
\hline
Star &  $V_{\star}$ & \multicolumn{1}{c}{$D$} &$T_{\mathrm{eff}}$& \multicolumn{2}{c}{$R_{\star}$}  \\
    &  (km s$^{-1}$) & \multicolumn{1}{c}{(pc)}&(K) & (mas) & (AU) \\
\hline
S Per  & --38.5 & $2312\pm^{65}_{32}$ &3550&\,\,\,$3.5\pm1.5$&8.0 \\
U Ori  & --39.5 & $260\pm50$  &2570& \,\,$5.7\pm0.2$&1.5 \\
U Her  & --14.5 & $266\pm^{32}_{28}$ \,\,\, &2630& \,\,$4.8\pm0.2$&1.3 \\
IK Tau & +34.0  & $250\pm20$ &--& $11.2\pm1.1$&2.8 \\
RT Vir & +18.2  & $135\pm15$ &2924 & \,\,$6.2\pm0.3$&0.8
\\
\multicolumn{6}{l}{References}\\
S Per  & D87 & M08 & L05 & \,\,H94, L05 &\\
U Ori  & C91 & \,\,C91 & A97 &\,\,R06  &\\
U Her  & C94 & \,\,V07 & A97 &\,\,R06  &\\
IK Tau & K87 & \,\,098 & -- &\,\,M04, R06  &\\
RT Vir & N86 & vL07 & C07 &\,\,M04  &\\

\hline
\end{tabular}
\caption{Properties of the sample stars. The stellar velocity
$V_{\star}$ is given in the Local Standard of Rest (LSR)
convention. The distances $D$ used in our calculations are 2.3 kpc for
S Per, 266 pc for the Miras, IK Tau, U Ori and U Her and 133 pc for
the SRb RT Vir, in order to remain consistent with RYC99 and B+03.
The values of $R_{\star}$ for the AGB stars were measured using IR
interferometry at H and K bands. The radius of S Per is the average of
values deduced from spectral fits. References in the table are A97 \citet{Alvarez97}; C91 \citet{Chapman91};
C94 \citet{Chapman94};C07 \citet{Cenarro07}; D87 \citet{Diamond87}; H74 \citet{Humphreys74};
K87 \citet{Kirrane87}; L05 \citet{Levesque05};
M04 \citet{Monnier04}; M08 \citet{Mayne08}; N86 \citet{Nyman86}; 098
\citet{Olofsson98}; R06 \citet{Ragland06}; V07 \citet{Vlemmings07}; vL07
\citet{vanLeeuwen07}.  }
\label{stars}
\end{table}

 \begin{figure*}
   \centering
   \includegraphics[angle=-90, width=15cm]{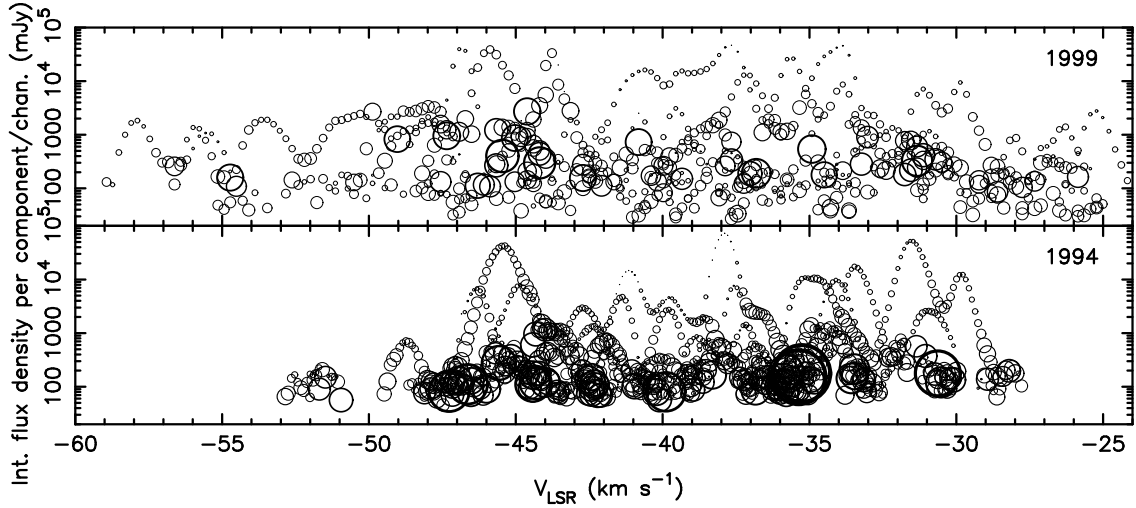}
      \caption{The components making up the velocity profiles of H$_{2}$O maser
features of S Per in 1999 and 1994. The symbol diameter is proportional to the
observed angular FWHM $s_{\sv}$ of the components.}
         \label{ALLFEATSSPER.PS}
   \end{figure*}

Observational parameters are given in Table~\ref{obstab}. The
$\sigma_{\mathrm{rms}}$ noise varies greatly not only due to the
different durations of observations but to the elevation, the peak
(determining the effectiveness of self calibration) and most of all
the weather.  The 1994
  observations and image analysis methods are described in detail by
  \citet{Bains03} (B+03) for the AGB stars IK Tau, RT Vir, U Her and U
  Ori and in \citet{Richards99} (RYC99) for the RSG S Per.  Similar
  procedures were used at later epochs, except that phase-reference
  sources were included after 1999 in order to improve position
  accuracy.  The 1994 observations were made with channels of 0.105 km
  s$^{-1}$ in a total span of 25 km s$^{-1}$, as was used for all
  epochs of U Ori, U Her and RT Vir.  The spectra of IK Tau and S Per
  were found to have a larger velocity extent, so double the span was
  used for these two sources in subsequent epochs, necessitating a
  doubling of the channel width to 0.21 km s$^{-1}$.  The image
  weightings and restoring beams were adopted to optimise the accuracy
  of component fitting, which depends mainly on the signal-to-noise
  ratio (RYC99) as long as the emission has a naturally Gaussian
  distribution. The masers are well-resolved by MERLIN; the
  total extent of the 22-GHz emission regions is over a hundred
  mas, at a resolution of 10--20 mas, and many spectral channels
  contain multiple distinct patches of maser emission.

RYC99 showed that we detect at least 85\% of the S Per 22-GHz maser emission.
Single-dish monitoring with the Pushchino telescope is available within a few
months of some epochs for some of our sources. \citet{Lekht05} detected
slightly less flux than we did for S Per in 1994, and more in 1999.
\citet{Rudnitskij00} show that we detected a similar amount of maser emission
from U Ori as they did in 1994 and 1999. The single-dish flux of RT Vir within
a few months of our observations (\citealt{Mendoza-Torres97};
\citealt{Lekht99}) was lower than the closest interferometry measurements, but
both sets of observations show that it is highly variable, changing flux
density by a factor of 2 in a few months.

\begin{table*}
\begin{tabular}{lcrccrrrrcrrrcrc}
\hline
Star & Date & \multicolumn{1}{c}{Dur}  & Bm & $\sigma_{\mathrm{rms}}$ &$r_{\mathrm{i}}$&$r_{\mathrm{o}}$&$\v_{\mathrm{i}}$&$\v_{\mathrm{o}}$&$\epsilon$& $NC$ &
$NF$ & $I_{\mathrm{max}}$ &$\overline{l}$ &$N_{\mathrm{fit}}$&$\overline{\Delta{V_{1/2}}}$\\
 & (yymmdd) & \multicolumn{1}{c}{(hr)}  &(mas)&(mJyb$^{-1}$)&
\multicolumn{2}{c}{(AU)}&\multicolumn{2}{c}{(km s$^{-1}$)}&&& &(Jy)
&(AU)&&(km s$^{-1}$) \\
\hline
S Per$^1$  &940324&14.5& 10 & 17 &55&175&9.0&16 &0.51 &1040& 93& 72&$18\pm9  $  &50&$0.77\pm0.29$\\
    S Per  &990110&13.0& 10 &\,\,\,8 &39&175&8.0&22 &0.56 & 689&100& 47&$12\pm6  $  &56&$0.91\pm0.34$\\
\\
    U Ori  &940417&12.6& 15 & 12 &10& 32&2.5&7.5&0.93 &  64& 14& 26&$2.7\pm1.8$& 2&$0.65\pm0.12$\\
    U Ori  &990109& 9.1& 20 & 20 &12& 36&2.5&5.5&0.72 &  94& 13& 44&$4.8\pm2.6$& 6&$0.51\pm0.16$\\
    U Ori  &000410& 8.2& 18 & 25 & 7& 36&2.0&6.0&0.65 & 142& 31&  8&$2.7\pm2.2$& 5&$0.46\pm0.23$\\
    U Ori  &010506& 5.8& 18 & 27 & 7& 29&2.0&5.5&0.68 & 165& 25& 10&$4.9\pm4.4$&10&$0.74\pm0.23$\\
\\
    U Her  &940413&13.7& 15 & 14 &13& 47&4.0&9.5&0.69 & 126& 34& 22&$2.3\pm1.7$&18&$0.68\pm0.35$\\
    U Her  &000519& 7.6& 18 & 40 &10& 41&3.0&8.0&0.72 & 303& 44&141&$4.9\pm4.4$&22&$0.71\pm0.27$\\
    U Her  &010427& 6.1& 18 & 35 &10& 41&3.0&8.0&0.72 & 212& 37& 38&$3.9\pm3.5$&16&$0.62\pm0.32$\\
\\
IK Tau$^1$ &940415&11.0& 15 & 10 &16& 66&5.0&16 &0.82 & 742&256& 25&$2.0\pm0.6$ &57&$0.68\pm0.28$\\
    IK Tau &000520& 7.3& 15 & 35 &16& 72&6.0&18 &0.73 & 407& 72& 84&$3.4\pm3.1$ &36&$0.98\pm0.41$\\
    IK Tau &010427& 3.9& 15 & 40 &16& 72&6.0&18 &0.73 & 175& 40& 24&$3.2\pm2.9$ &19&$0.81\pm0.31$\\
\\
    RT Vir &940816&11.8& 20 & 12 & 6& 25&4.0&10 &0.65 & 581& 55&394&$1.4\pm0.6$ &40&$1.04\pm0.38$\\
    RT Vir &960405& 4.5& 12 & 25 & 5& 19&3.5&11 &0.81 & 389& 58&389&$1.2\pm0.9$ &21&$0.87\pm0.37$\\
    RT Vir &960421&10.8& 12 & 30 & 5& 19&3.5&11 &0.81 & 400& 52&516&$1.2\pm0.9$ &20&$0.90\pm0.36$\\
    RT Vir &960429&11.5& 12 & 15 & 5& 19&3.5&11 &0.81 & 450& 50&526&$0.9\pm0.6$ &23&$0.97\pm0.41$\\
    RT Vir &960515& 9.5& 12 & 25 & 5& 19&3.5&11 &0.81 & 433& 41&727&$1.0\pm0.8$ &20&$1.05\pm0.42$\\
    RT Vir &960524& 7.7& 12 & 20 & 5& 19&3.5&11 &0.81 & 401& 51&706&$0.9\pm0.7$ &19&$0.88\pm0.31$\\
    RT Vir &960612& 8.4& 12 & 35 & 5& 19&3.5&11 &0.81 & 246& 38&791&$0.9\pm0.6$ &19&$0.80\pm0.25$\\
\hline
\end{tabular}

\caption{The first 5 columns give the target, date, duration,
resolution and sensitivity of each observation.   $r_{\mathrm{i}}$, $r_{\mathrm{o}}$, $\v_{\mathrm{i}}$
and $\v_{\mathrm{o}}$ are the inner and outer limits of the H$_{2}$O
maser shell.  $\epsilon$ is the logarithmic velocity gradient
(Section~\ref{clouds:denser}). The total number of individual maser
components ($NC$) and features ($NF$) made up of contiguous series of
components are given, along with the maximum total intensity
$I_{\mathrm{max}}$ of the brightest spatially distinct
component. $\overline{l}$ is the mean size of each feature,
$N_{\mathrm{fit}}$ is the number for which a Gaussian spectral profile
could be fitted at $>3\sigma$ and $\overline{\Delta{V_{1/2}}}$ is the
mean FWHM of these. All observations used 0.105 km s$^{-1}$ channel
width apart from S Per epoch 990110 and IK Tau epochs 000520 and
010427, when 0.21 km s$^{-1}$ was used.
\newline $^1$$NC$, $NF$
and $\v_{\mathrm{o}}$ are probably underestimates since the observational
bandwidth was less than the likely velocity extent of emission.
} \label{obstab}
\end{table*}

\subsection{Analysis of maser images}
\label{sec:images}
Detailed results from the 1994 observations are given in RYC99 and B+03 and the
large-scale behaviour revealed by more recent observations will be analysed in
a future paper.  Here, we concentrate on the properties relevant to
understanding maser beaming, which are summarised in Table~\ref{obstab}.

Identifying individual velocity-coherent maser features in an
expanding, turbulent outflow is a difficult task. To this end we
developed a two-stage process to disentangle the maser structure
properly in velocity and space, first by characterising the spatial
distribution of the emission channel by channel and then by examining
the spatially-resolved spectral profiles.  We start by introducing the
first concept, of a {\em maser component}: Emission measured in a
particular channel centred on velocity \v, in a contiguous area on the
plane of the sky centred on a given direction.  The emission pattern
of a 3D unsaturated maser in a given frequency interval has the
Gaussian shape $e^{-(\theta/\theta_{\sv})^2}$, where $\theta$ is
displacement from the direction of the longest chord along the
line-of-sight and $\theta_{\sv}$ is the beaming angle; the observed
area is then $A = \pi\theta_{\sv}^2$ \citep{Elitzur92}. This property
of maser emission allows us to obtain accurate measurements by fitting
a two-dimensional Gaussian component to each patch of maser emission, in
each channel, and deconvolving the restoring beam from the map
area as described in RYC99. We selected components brighter than 4 or
5$\sigma_{\mathrm{rms}}$ noise (see RYC99 and B+03 for more details of
rejection of artefacts).  The resulting fit determines the intensity
$I$, position ($x$, $y$) and angular FWHM (full width at half maximum)
$s_{\sv}$ ($= 2\sqrt{\ln2}\,\theta_{\sv}$) of each maser component. 

The uncertainty $\sigma_s$ in the component size is the sum in quadrature of
the position uncertainties in the $x$ and $y$ directions.  There is a lower
limit of 0.1 mas to the resolvable component size \citep{Richards97t} due to
dynamic range and calibration limitations. Between 1\% and 40\% of component
sizes were upper limits and if $\sigma_s> s_{\sv}$, we took $s_{\sv}=\sigma_s$.
Figure~\ref{SPOTHIST.PS} shows the distribution of maser component sizes
(restricted to $s_{\sv}>\sigma_s$) for each source and epoch. The brightness
temperature of a component is
\begin{equation}
    T_{\mathrm {B}} = \frac{10^{-26} I \lambda^2}{2k_{\mathrm{B}} A}
 \label{Tb}
\end{equation}
where $I$ is in Jy, the wavelength $\lambda=0.013$ m, $k_{\mathrm{B}}$ is
Boltzmann's constant and the area $A = (\pi/4\ln{2})s_{\sv}^2$ is in steradian.


 \begin{figure}
   \centering
   \includegraphics[angle=-90, width=9cm]{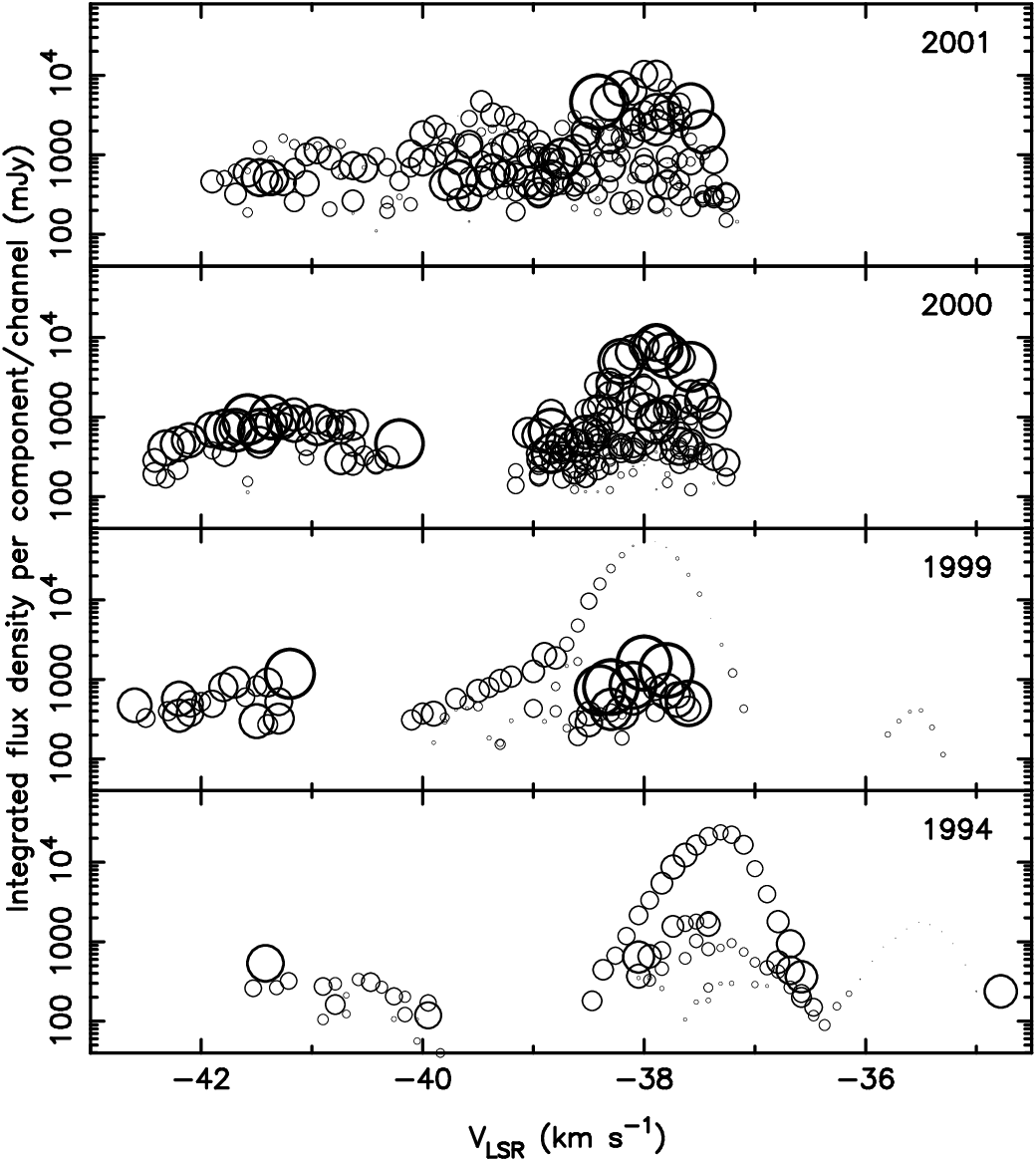}
      \caption{The velocity profile of U Ori, see Fig.~\ref{ALLFEATSSPER.PS} for details.}
         \label{UOSPVELPROF.PS}
  \end{figure}
 \begin{figure}   \centering
   \includegraphics[angle=-90, width=9cm]{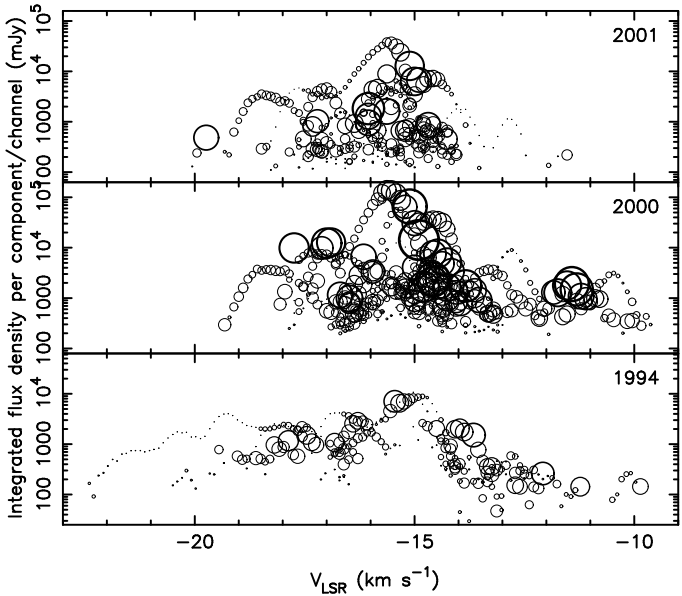}
      \caption{The velocity profile of U Her, see Fig.~\ref{ALLFEATSSPER.PS} for details.}
         \label{UHSPVELPROF.PS}
   \end{figure}
\begin{figure}
   \centering
   \includegraphics[angle=-90, width=9cm]{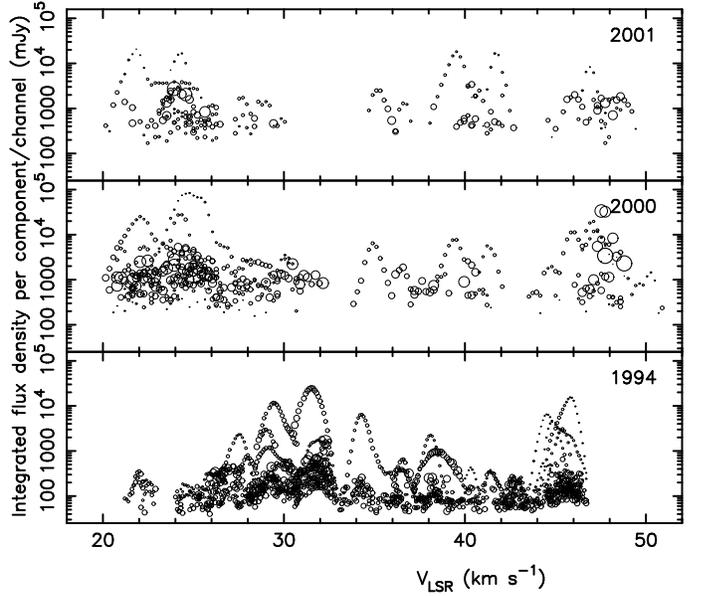}
      \caption{The velocity profile of IK Tau, see Fig.~\ref{ALLFEATSSPER.PS} for details.}
         \label{IKSPVELPROF.PS}
   \end{figure}

\begin{figure}
   \centering
   \includegraphics[angle=-90, width=9cm]{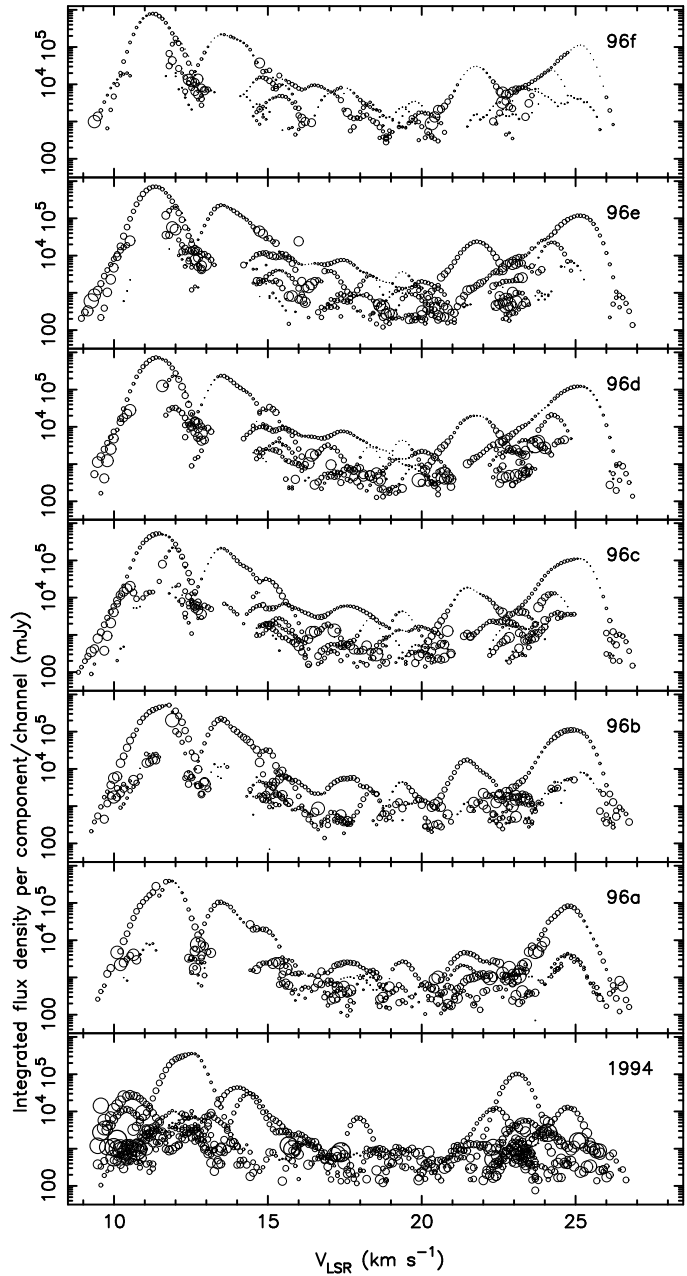}

\caption{The velocity profile of RT Vir, see Fig.~\ref{ALLFEATSSPER.PS} for
details. 96a to f indicate the six successive epochs of observation in 1996.}
        \label{RTSPVELPROF.PS}
   \end{figure}

   \begin{figure}
   \flushleft   
\includegraphics[angle=0, width=9.3cm]{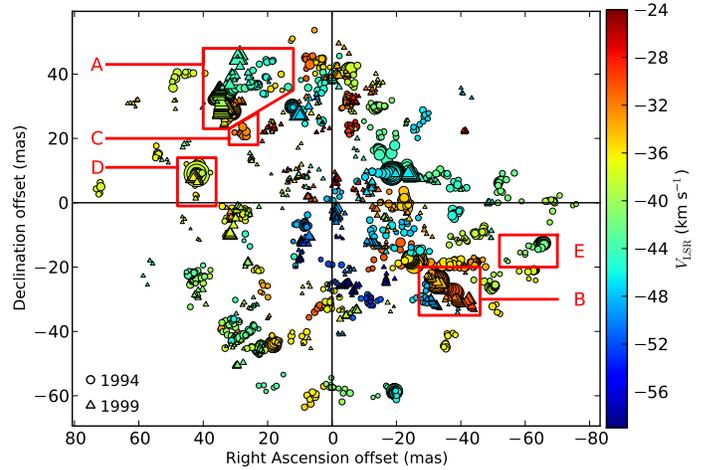}
      \caption{Each symbol represents an H$_{2}$O maser component in a
              single channel observed towards S Per in 1994 ($\circ$)
              or 1999 ($\triangle$). The
              diameter is proportional to the square root of total
              intensity. The colour is proportional to the velocity as
              shown.  The red boxes enclose features which are shown
              in more detail in Figs.~\ref{F58.PS}, ~\ref{94POS.PS}
              and~\ref{MULTIFEATSS.PS}.  }
         \label{SPER_9499_BOXES.PS}
   \end{figure}

\begin{figure}
 \centering
   \includegraphics[angle=0, width=9cm]{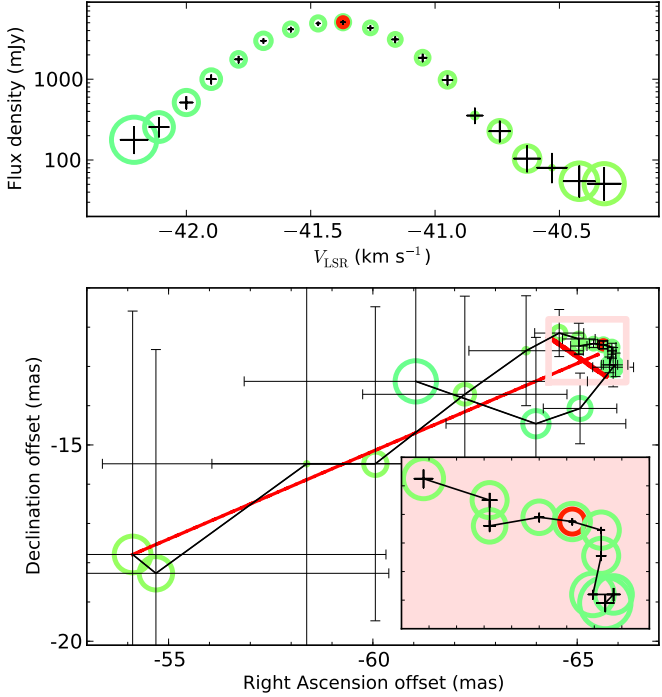}
\caption{Properties of the maser feature {\sf E} in
Fig.~\protect{\ref{SPER_9499_BOXES.PS}}, with the same coloring
scheme.  {\em Top}: Feature line profile. The symbol diameters are
proportional to $s_{\sv}$ (two are very small); the error bars show
the size uncertainty $\sigma_s$.  {\em Bottom}: Detailed map of the
components comprising the feature. The diameter of each symbol is 10\%
of $s_{\sv}$ and the error bars represent the position offset
uncertainties. The thin black line joins the components in velocity
order. The long thin red line denotes the largest angular size $l$ of
the feature. The short fat red line shows the separation of the
components with flux densities closest to the half-maximum of the
feature, $d$. The expanded-scale inset (total size $1.7\times1.2$ mas,
ticks at 0.2 mas intervals) shows the area in the pink box, which
contains the components brighter than half maximum, with component
size error bars. The brightest component is outlined in red, also
marked on the {\em Top} plot. }
         \label{F58.PS}
   \end{figure}

\begin{figure}
 \centering
   \includegraphics[angle=0, width=9cm]{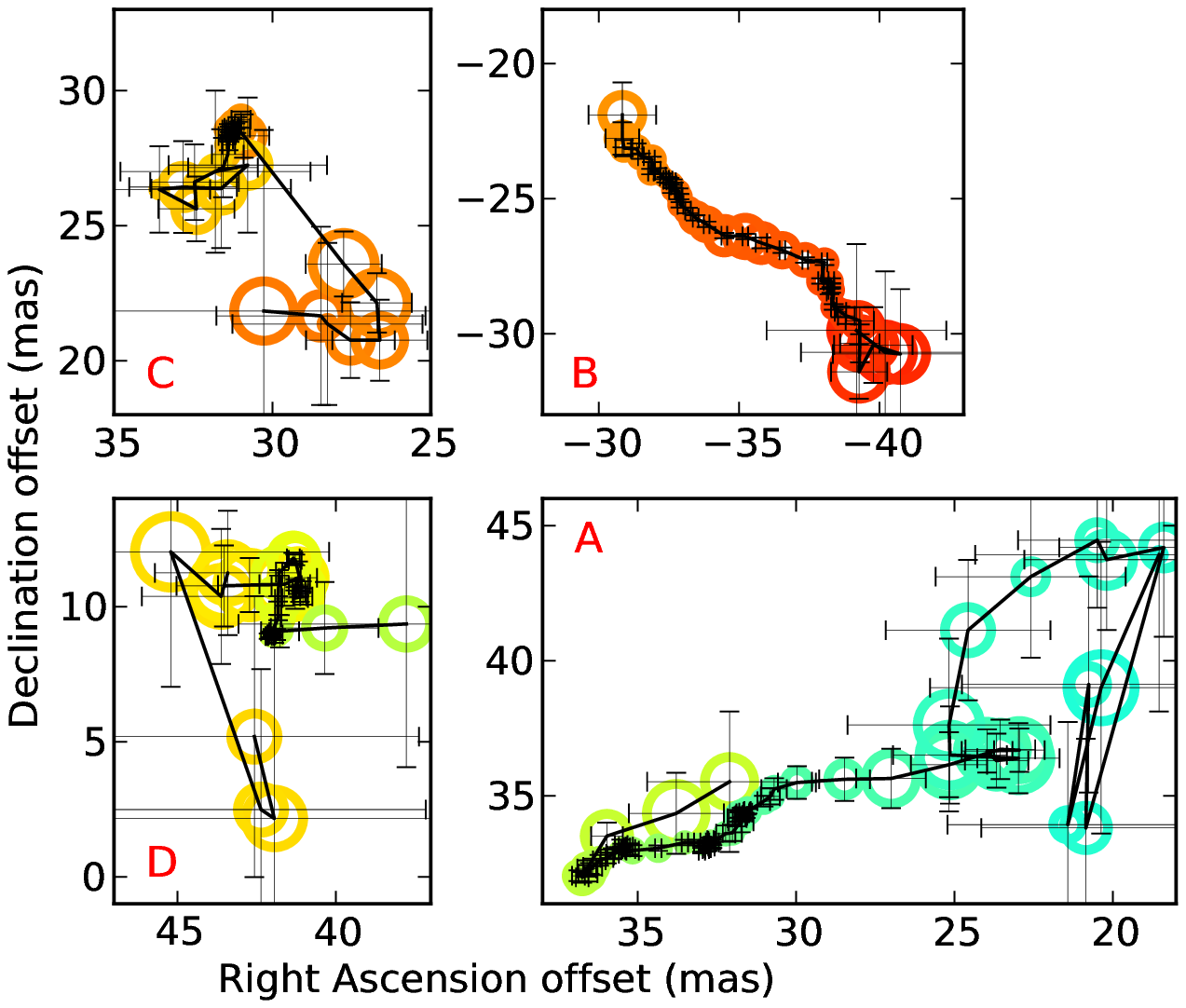}
   \includegraphics[angle=0, width=4.5cm]{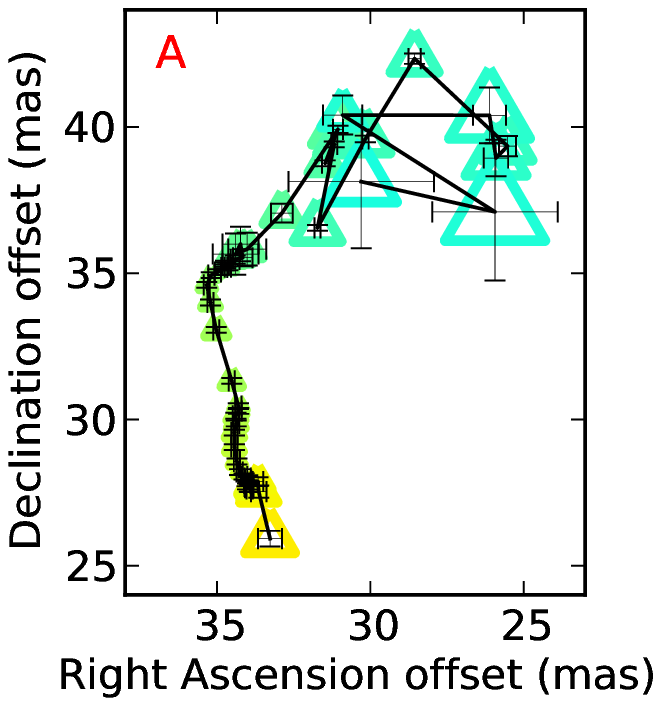}

\caption{The angular size, distribution and position uncertainties of
the maser components making up the boxed features {\sf A}--{\sf D} around S Per in
Fig.~\protect{\ref{SPER_9499_BOXES.PS}}, as measured in 1994 (upper
and middle plots, circles) and the features in {\sf A} as measured in
1999 (lower plot, triangles). The same velocity colour scheme is
used. Region A contains 6 features in 1994 and 5 in 1999; the other
regions contain 2 features, as can be seen from the spectral peaks in
Fig.~\ref{MULTIFEATSS.PS}. The diameter of each symbol is 10\% of
$s_{\sv}$.  The black lines join the components in velocity order. The
position uncertainties are shown.
 }
         \label{94POS.PS}
\end{figure}

   \begin{figure}
   \centering
   \includegraphics[angle=-90, width=8cm]{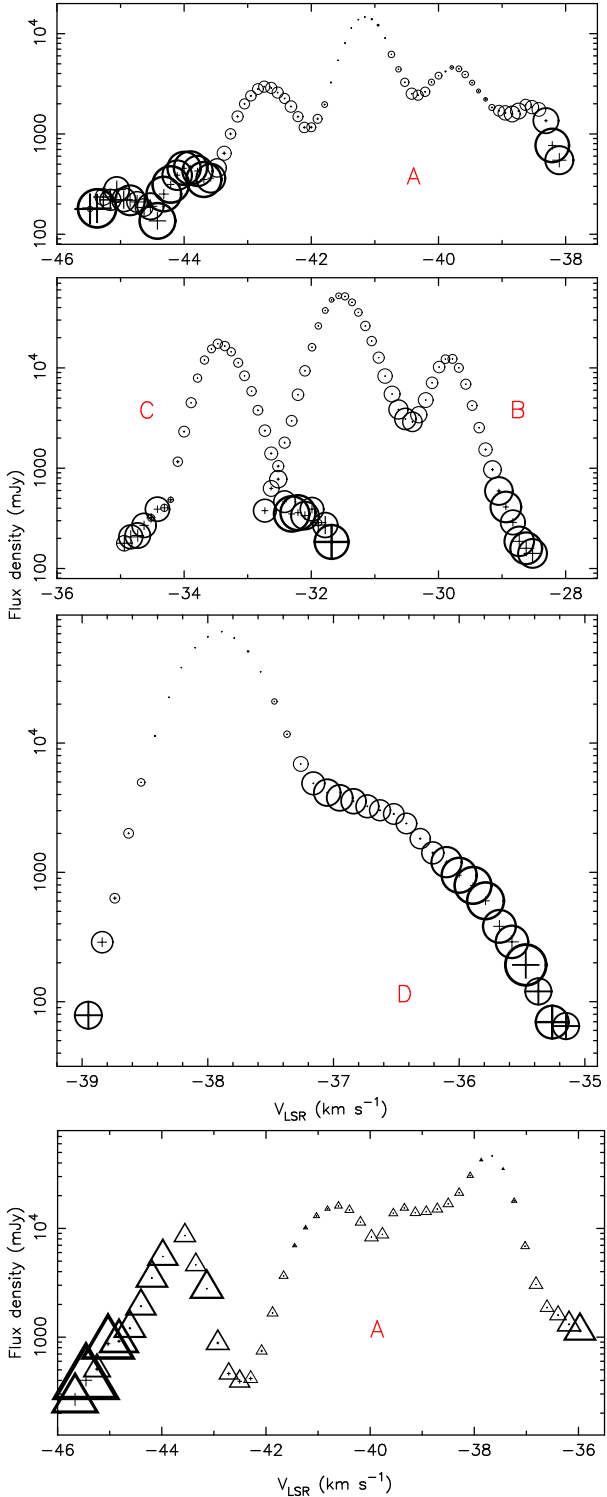}
   \includegraphics[angle=-90, width=8cm]{REYMaserSpotv3_f13.ps}
      \caption{Top to bottom: The line profiles of features seen
  around S Per in 1994 (circles) in the regions labelled {\sf A}, {\sf
  B}, {\sf C} and {\sf D} in Fig.~\protect{\ref{SPER_9499_BOXES.PS}},
  and in 1999 (triangles) in the region labelled {\sf A} in
  Fig.~\protect{\ref{SPER_9499_BOXES.PS}}.  The diameters of the
  circles/triangles and of the crosses are proportional to the observed angular FWHM
  $s_{\sv}$ of the maser components and to $\sigma_{s}$, respectively.
  }
         \label{MULTIFEATSS.PS}
  \end{figure}

  \begin{figure}
   \flushleft
   \includegraphics[angle=0, width=9.5cm]{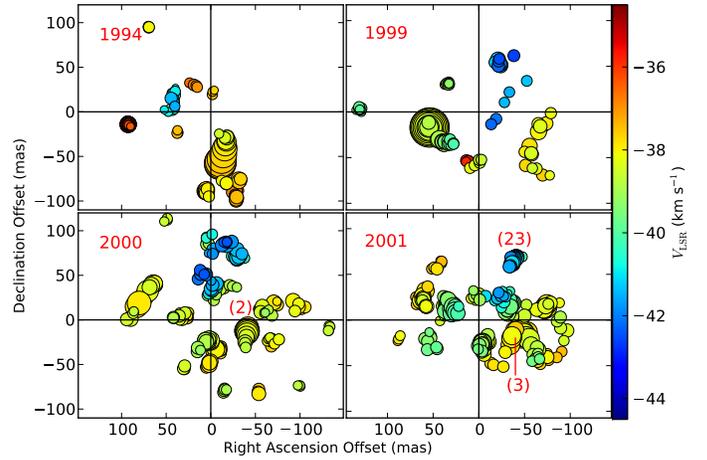}
      \caption{Each symbol represents an H$_{2}$O maser component in a
              single channel observed towards U Ori, at the 4 epochs
              indicated.   The
              diameter is proportional to the square root of total
              intensity. The colour is proportional to the velocity as
              shown.
               The numbered features are shown in more detail in
              Figs.~\ref{UORISPOTPOS.PS} and~\ref{UOSELSPVELPROF.PS}.
}
         \label{UORIALL000_FF.PS}
   \end{figure}

\begin{figure}
 \centering
   \includegraphics[angle=0, width=9cm]{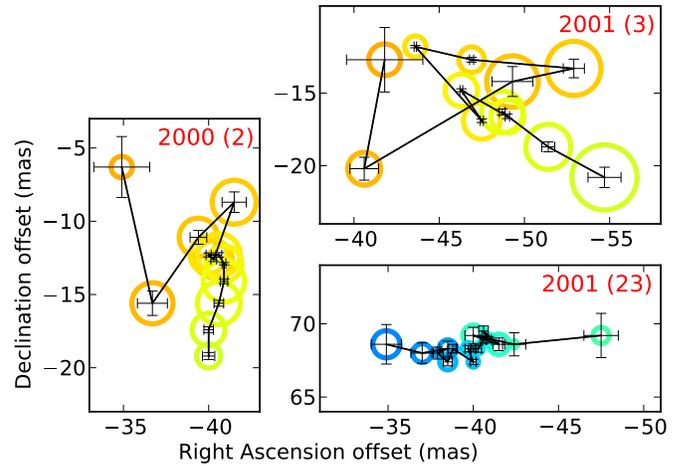}

\caption{The angular size and distribution of the maser components in the U Ori
features marked on Fig.~\ref{UORIALL000_FF.PS}, with the same
colouring scheme. Each panel shows a single feature. The diameter
of each symbol is 10\% of  $s_{\sv}$ and the error bars represent the position
offset uncertainties. The line joining the components shows the velocity order.
}
         \label{UORISPOTPOS.PS}
\end{figure}
\begin{figure}
   \centering
   \includegraphics[angle=-90, width=9cm]{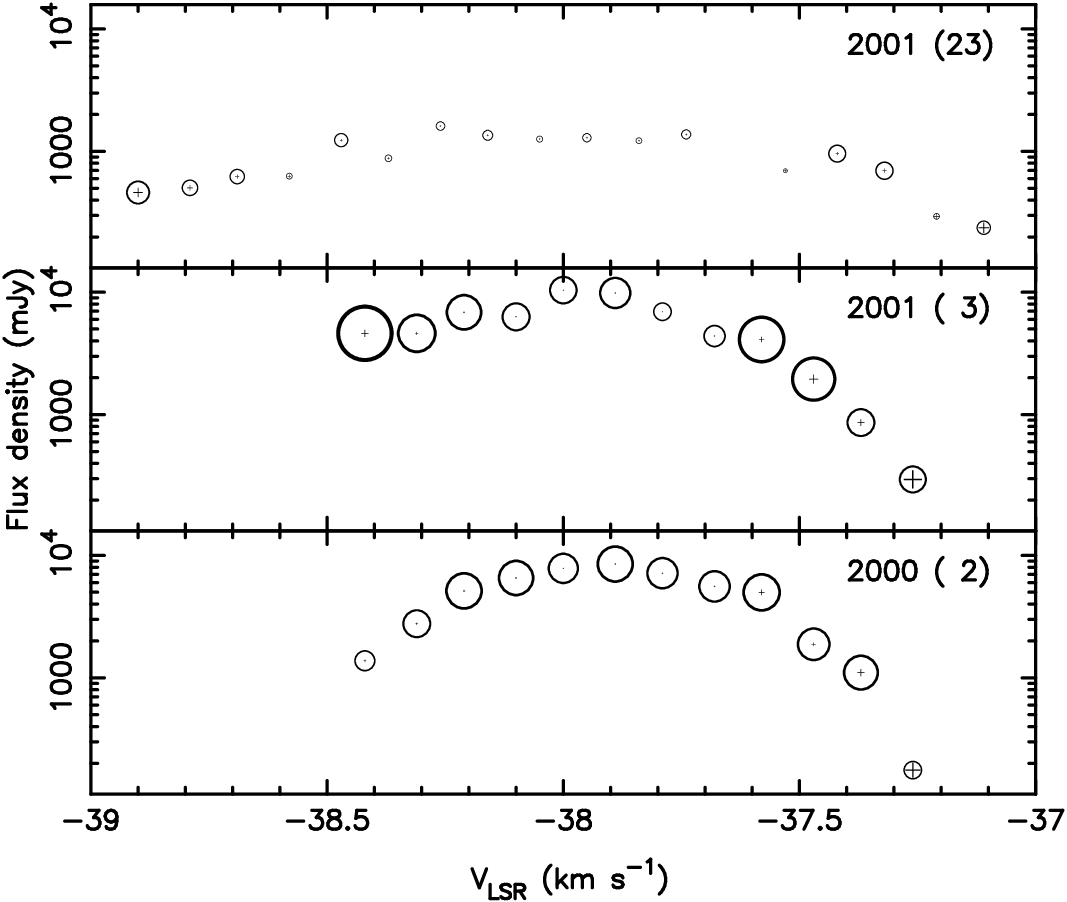}

\caption{The line profiles of U Ori features marked on
Fig.~\ref{UORIALL000_FF.PS}.   The diameters of the circles and crosses are
proportional to the observed angular FWHM $s_{\sv}$ of the maser components and
to $\sigma_{s}$, respectively.
}
         \label{UOSELSPVELPROF.PS}
\end{figure}


Figs.~\ref{ALLFEATSSPER.PS}--\ref{RTSPVELPROF.PS} show the variations of
$s_{\sv}$ across the line profiles at each epoch.
Figs.~\ref{SPER_9499_BOXES.PS}, \ref{UORIALL000_FF.PS} and \ref{RTALLPOS.PS}
illustrate the component distributions in the most well-filled, symmetric CSE
(S Per), the most sparse (U Ori) and a well-filled CSE with velocity asymmetry
(RT Vir); see plot captions for details. These Figures show that maser
components lie in series like pearls on a necklace, often curved or
twisted.

We
introduce the second concept, of a {\em maser feature}: A series of adjacent components
within a beam width, spanning at least 3 velocity channels.  Each feature was
defined as possessing a single-peaked spectral profile, unless the emission was
too faint to show a clear peak or the profile was truncated by blending. All components were allocated to features
consisting of collections of neighbours within a beam width in successive
channels.  In cases of ambiguity, component allocation was guided by following
the smoothest increments of position and flux density from channel to channel.
A few, isolated components were rejected as artefacts, being very faint and/or
in the position of beam side lobes.

 An example feature is enlarged in Fig.~\ref{F58.PS} (region {\sf E}
in Fig.~\ref{SPER_9499_BOXES.PS}). This shows that the smooth line
profile corresponds to a curved position-velocity gradient.  The
angular FWHM $d$ of the feature is given by the separation of
components with intensities closest to half maximum intensity, marked
by the short fat red line. The largest angular size $l$ (uncertainty
$\sigma_l$) between its most widely separated components is shown by
the long red line in Fig.~\ref{F58.PS}. The interpretation of $d$ is
discussed further in Section~\ref{saturation:beaming}.

We consider the features thus identified as the fundamental building
blocks of the maser source.  We fitted the spectral profile of each
feature, i.e.  the component intensity $I(\v)$ as a function of \v,
with a Gaussian curve.  This is an approximation to the line shape of
unsaturated masers, adequate to half maximum.  This fitting yields the
line FWHM $\Delta{V_{1/2}}$, and the corresponding uncertainty
$\sigma_{\Delta{V_{1/2}}}$, as well as the central velocity, the peak
model intensity and their uncertainties. The relevant measurements are
given in Table~\ref{obstab}, along with the number of features at each
epoch with fits better than $3_{\sigma}$. 

We show additional enlargements of features in
Figs.~\ref{94POS.PS},~\ref{MULTIFEATSS.PS}, \ref{UORISPOTPOS.PS},
\ref{UOSELSPVELPROF.PS}, \ref{RTSPOTPOS.PS} and \ref{RTSELSPVELPROF.PS} as
clear (but otherwise unexceptional) illustrations of different relationships
between $s_{\sv}$ and intensity or position in the line profile, discussed in
Section~\ref{beaming}.  Figs.~\ref{94POS.PS} and
\ref{MULTIFEATSS.PS} show multiple spectral peaks in spatially adjacent
features which probably emanate from a single physical region, discussed in
Section~\ref{clouds:profiles}.

In summary, the individual maser \emph{components} represent separate patches
of emission as sampled in the observed velocity channels.  The components are
found to be clustered in position and velocity, often forming smooth Gaussian
spectral profiles, termed \emph{features}. Individual features thus defined are
the objects to be compared with maser theory. Table~\ref{obstab} gives the
number of components and features in each source.

\section{Properties of maser clouds}
\label{cloudproperties}
\subsection{Dense clumps in the stellar winds}
\label{clouds}

The evidence for the origins of 22-GHz H$_2$O maser emission in
discrete, density-bounded clouds, as part of a fragmented, expanding
wind, is summarised in this section. 
  RYC99 and B+03 explained in more detail why we think that each feature (or
occasionally a series of a few features, in the RSG) corresponds to a
physically discrete, density bounded \emph{cloud}, not solely a
`spooky' accident of the velocity field or turbulence
\citep{Strelnitski07}.

\subsubsection{Maser components form distinct features with Gaussian
  spectral profiles}
\label{clouds:profiles}
The relationship between component positions and velocities is
illustrated for S Per (Figs.~\ref{SPER_9499_BOXES.PS}
and~\ref{94POS.PS}), U Ori (Figs.~\ref{UORIALL000_FF.PS} and
\ref{UORISPOTPOS.PS}) and RT Vir (Figs.~\ref{RTALLPOS.PS}
and~\ref{RTSPOTPOS.PS}).  Many features show an orderly (but not
necessarily linear) distribution of components, with an internal
gradient of velocity with position.  These gradients do not show any
large-scale alignment.
The average line widths $\Delta{V}_{1/2}$ for the feature spectral
profiles for each star are given in Table~\ref{obstab}.  The gas
temperature $T$ in the CSE of AGB stars is likely to drop from around
1000 K at $r_{\mathrm{i}}$ to 400 K at $r_{\mathrm{o}}$, depending on
the thickness of the shell \citep{Zubko00}. 

The temperature at
distance $r$ from the star can be approximated as $T\propto r^{-0.4}$,
so the thermal Doppler line width $\Delta{V}_{\mathrm{D}} \propto
\sqrt{T} \propto r^{-0.2}$.  $\Delta{V}_{\mathrm{D}}$ lies in the
range 1.6--1.2 km~s$^{-1}$ between $r_{\mathrm{i}}$ and
$r_{\mathrm{o}}$. The weak temperature dependence means that even if
RSGs and their CSEs are slightly hotter than AGB stars (Table
~\ref{stars}), this has little effect on maser amplification paths.
$\Delta{V}_{\mathrm{D}}$ exceeds $\Delta{V}_{1/2}$, so maser
amplification is likely to take place through the entire depth of each
feature.

Features around AGB stars are almost always distinct in position or
velocity (or both) and we conclude that they correspond to separate
clouds.  A few features around
RSGs  form groups such as the example in
Fig.~\ref{SPER_9499_BOXES.PS} region {\sf A}. This appears to be a single
cloud containing 5--6 multiple, adjacent spectral peaks
(Fig.~\ref{MULTIFEATSS.PS}), i.e. multiple features in our
terminology.  These large clouds have a velocity span
$>\Delta{V}_{\mathrm{D}}$ and RYC99 showed that their velocity
gradients, rather than their total depths, probably limits their maser
amplification.  The  majority (80\% -- 90\%) of features around RSGs
appear to emanate from separate clouds.

\subsubsection{The angular extent of features gives the clouds
  physical sizes}
\label{clouds:sizes}

Emission from successive velocity channels sampling a quiescent sphere
would be centred on a common point on the sky. Alternatively, a long
thin line of observed maser components would be produced by a
spherical cloud with a linear internal velocity gradient.  The
patterns we observe could be produced by spherical clouds with curved
velocity gradients.  It seems likely that MERLIN is detecting all the
maser emission (Section~\ref{sub:obs}) so $l$ is close to the full
extent of the maser features with significant position-velocity
gradients.  In other cases, $l$ is a lower limit, but individual
clouds cannot exceed about 20 $\overline{l}$ in any direction or they
would not fit inside the observed maser shell inner and outer radii.

The clouds are not necessarily spherical but could be amoeboid or have
any other shape. Unfortunately we cannot measure all three dimensions
of each cloud directly so we use $l$ (converted to units of length at
the appropriate distance for each star) as the physical diameter of a
feature in any direction, indicating the most likely biases this
causes where relevant.  The analysis in this paper (e.g.
Section~\ref{discussion}) will provide further clues to the shapes of
clouds.

\subsubsection{Maser clouds persist for many months or years}
\label{clouds:persist}
 The distribution of components within some features forms distinctive
patterns which can be recognised (with some distortion) at multiple
epochs.  Close similarities can be seen in some instances, such as the
RT Vir feature shown in Fig.~\ref{RTSPOTPOS.PS} (one of 11 features
matched at all 6 epochs of observation taken over 10 weeks). The S Per
features seen in Fig.~\ref{SPER_9499_BOXES.PS} region {\sf A}, also
shown in Fig.~\ref{94POS.PS}, appear to have rotated about 60\degr\/
anticlockwise during 5 years.  The epochs were aligned using the
centres of expansion (found as described in RYC99), combined, relative
uncertainties $\sim6.5$ mas.  Proper motion studies of the AGB stars
show that masing from their H$_{2}$O clouds can be tracked for between
a few months and a couple of years (\citealt{Richards99m};
\citealt{Yates94}).  The larger RSG clouds can be tracked for 5 to 10
or more years (\citealt{Richards96}; \citealt{Richards98V};
\citealt{Murakawa03}).

\subsubsection{Water maser clouds are much denser than their surroundings}
\label{clouds:denser}
The H$_{2}$O masers propagate in approximately spherical shells with
inner and outer radii of $r_{\mathrm{i}}$ and $r_{\mathrm{o}}$, across
which the expansion velocity increases from $\v_{\mathrm{i}}$ to
$\v_{\mathrm{o}}$. Multi-epoch and proper motion measurements suggest
that the H$_{2}$O maser shells are approximately spherically symmetric
(\citealt{Murakawa03}, \citealt{Richards99m}).  This
allowed \citet{Murakawa03} (their equation 1) to solve the
three-dimensional structure of the circumstellar envelopes and find
the distance of each cloud from the star, $r$.

The wind acceleration through the shell is
parametrised by the logarithmic velocity gradient
 $   \epsilon = \log(\v_{\mathrm{o}}/\v_{\mathrm{i}})/
               \log(r_{\mathrm{o}}/r_{\mathrm{i}})
$. 
\citet{Alcock86} proposed that if initially spherical, dusty clouds
  are driven away from a star by radiation pressure on grains, at a
  constant velocity, their diameter in the radial, outflow direction
  will be unchanged but their diameter in the tangential direction
  will increase in proportion to distance from the star.  If the wind
  is accelerating, then the tangential and radial axes (with respect
  to the star) of a cloud expanding under radiation pressure will grow
  as $r/r_{\mathrm{i}}$ and $(r/r_{\mathrm{i}})^{\epsilon}$,
  respectively. If the clouds are spherical at $r_{\mathrm{i}}$, their
  tangential/radial aspect ratio will evolve as
  $(r/r_{\mathrm{i}})^{1-\epsilon}$.  The cloud number density,
  relative to that at the inner radius, is given by
  $n(r)=n_{r_{\mathrm{i}}}\times (r/r_{\mathrm{i}})^{-(2+\epsilon)}$.

The inner radius $r_{\mathrm{i}}$ is determined by the distance from the star
where the wind density falls below the collisional quenching density for the
22-GHz H$_{2}$O maser (\citealt{Cooke85}; \citealt{Cohen87a}), at a number
density of $n(r_{\mathrm{i}})\approx 5\times10^{15}$ m$^{-3}$
($\approx 5\times10^{9}$ cm$^{-3}$).
The mass loss rate $\dot{M}$ which would be needed to
produce $n(r_{\mathrm{i}})$ is one or two orders of magnitude greater than the
mass loss rates measured from IR and CO observations (see references in RYC99
and B+03). 

 This apparent contradiction is overcome by considering the volume
filling factors of the H$_{2}$O maser clouds. The number of clouds
$NC$ and their sizes $l$ are given in Table~\ref{obstab} and the
filling factors are given by $NC\times
l^3/(r_{\mathrm{o}}^3-r_{\mathrm{i}}^3)$. S Per has the highest factor
of 1.5\% (RYC99), RT Vir and IK Tau have factors of about 0.4\% and U
Ori and U Her have factors of about 0.1\% (B+03). This leads to
realistic values of $\dot{M}$ if the clouds are surrounded by gas at
$\sim1/50$ of the cloud number density.
This implies that the actual average feature volume per CSE must be no
more than a factor of about two greater than that based on the average
measured $\overline{l}$, or the mass in features would exceed the total
mass loss rate.

\subsection{Maser optical depth and saturation state}
\label{saturation}

In this Section,
we use our measurements to provide estimates of the parameters
determining the maser optical depths, which leads us to infer that the
masers are mostly unsaturated.  We provide estimates of the
uncertainties in our measurements and the underlying assumptions;
although these are large, different approaches give consistent
results.

\subsubsection{Estimating the population inversion}
\label{saturation:inversion}

$\Delta{n} = n_2 - n_1$ is the difference between the number densities
per substate of the upper and lower levels involved in the maser
transition, $6_{16}$ and $5_{23}$ of ortho-water
\citep{deJong73}. Maser amplification requires $\Delta{n}>0$.

We start by assuming that the emission rate is approaching the pump rate
but not limited by it, following the assumptions outlined in 
RYC99 (section 4.5 and
references therein), which explains why this is reasonable, at least
for the brighter masers.
The population difference is given by equation
6.32 of \citet{Reid88}.   RYC99 (their equation 16) derived an
approximation in terms of measurable quantities; omitting constants
this is:
\begin{equation}
\Delta{n}\propto\frac{\Delta{V_{1/2}}T_{\mathrm{B}}}{\Gamma \lambda^2 l}
\label{Dn}
\end{equation}
for spherical clouds.
 $\Gamma$ is the decay rate from the
excited state, initially taken as 1 s$^{-1}$ \citep{Reid88}.

 We obtained mean values of $10^3 \le
\overline{\Delta{n}} \le 4\times10^4$ m$^{-3}$ ($0.001 \le
\overline{\Delta{n}} \le 0.04$ cm$^{-3}$) for S Per, U Ori, U Her and IK
Tau.  RT Vir had consistently high values, $10^4 \le
\overline{\Delta{n}} \le 9\times10^5$ m$^{-3}$ ($0.01 \le
\overline{\Delta{n}} \le 0.9$ cm$^{-3}$).

The population of the lower state $n_1$ was estimated using
$(n_1/n_{\mathrm{H_2O}})(n_{\mathrm{H_2O}}/n_{\mathrm{H_2}})n_{\mathrm{H_2}}$.
We took the fractional number density of H$_2$O
with respect to H$_2$ ($n_{\mathrm{H_2O}}/n_{\mathrm{H_2}}$) as
$2\times10^{-4}$, where  the fraction of H$_2$O molecules in the $n_1$ state is
$5\times10^{-3}$.  The actual values are likely to differ by a factor
of 2 or more, both between and within sources.  We take
$n_{\mathrm{H_2}}$ as
65\% of the total number density $n$, and estimate $n$ as a function
of $r$ as outlined in  Section~\ref{clouds:denser}.

If the pump rate is much higher than the stimulated emission rate,
$\Delta{n}$ is a lower limit.  If the masers are significantly
saturated then the present assumptions about the geometry are affected
and $\Delta{n}$ could be overestimated.  Errors in $r$ are unlikely to
exceed 30\%, producing less than a factor of $\sim$2 uncertainty in
$n$, and hence $n_1$. $\Delta{n}$ will be overestimated if $\Gamma$ or
the cloud depth are underestimated.  The average cloud depth is
unlikely to be more than about twice $\overline{l}$ and no individual
cloud depth can exceed $\sim20l$, as explained in
Section~\ref{clouds:denser}.  $\Gamma$ is unlikely to be that much
greater than our assumed value, so $\Delta{n}$ is very unlikely to be
more than a factor of 20 too large.

The mean percentage population inversion (weighted by measurement
errors), $100\times\Delta{n}/(n_1+n_2)$, was $\la1\%$ for all sources
apart from five RT Vir epochs and one IK Tau epoch which reached
between $1\%$ and $\sim10$\%.  Selecting brighter features, with the
brightness temperature at line peak
$T_{\mathrm{B0}}>10^9$ K, gave a higher mean population inversion of a
few percent for S Per, U Ori and U Her at most epochs with meaningful
data and a few tens of percent for one IK Tau epoch and RT Vir at most
epochs.

\subsubsection{Maser excitation temperature}
\label{saturation:Tx}
The maser excitation temperature is defined as
\begin{equation}
T_{\mathrm{x}} = (h\nu/k_{\mathrm{B}})/\ln[n_1/(\Delta{n}+n_1)]
\label{Tx}
\end{equation}
where $h$ and $k_{\mathrm{B}}$ are Planck's and Boltzmann's constants
respectively. For $v=22.235$ GHz, $h\nu/k_{\mathrm{B}}\approx1$ K.
We derived $T_{\mathrm{x}}$ at each
feature peak using our values of $\Delta{n}$ and $n_1$.  The
error-weighted mean $\overline{T_{\mathrm{x}}}$ for each epoch (given
in Table~\ref{alpha}) was --5 to  --0.2 K. S Per, U Her and RT Vir
have $\overline{T_{\mathrm{x}}}\ga-1$ K, as expected for the
consistently brighter sources.
The logarithmic relationship means that even if  $\Delta{n}$ is
decreased by a factor of 20, most
of the values of $\overline{T_{\mathrm{x}}}$ are decreased to not less
than --20 K, apart from for U Ori where it approaches --100 K,
although based on few significant measurements.

\subsubsection{Maser optical depth}
\label{saturation:tau}
The maser optical depth for the brightest part of each cloud is
related to the brightness temperature by
\begin{equation}
\tau=\ln(T_{\mathrm {B}0}/|T_{\mathrm{x}}|).
\label{tau}
\end{equation}
The mean ($\overline{T_{\mathrm {B}0}}$) and extrema
of $T_{\mathrm {B}0}$ for each source are given in
Table~\ref{alpha}. The maxima may be underestimated, if the brightest
peak component size is less than the minimum resolvable size. The
numbers of features, $NF$, detected at each epoch (Table~\ref{obstab}),
 do not have any significant relationship with
$3\sigma_{\rm{rms}}$ (except possibly for IK Tau), suggesting that the
minima are not  sensitivity-limited.

 We derived $\tau$ using our estimated
values of $T_{\mathrm{x}}$.  Table~\ref{alpha} gives the mean values
$\overline{\tau}$ for each epoch. The dispersion includes the scatter
of measured values but not the uncertainties in the underlying
assumptions.  $\overline{\tau}$ lies between 8 and 16 for S Per, U
Her, IK Tau and the epochs of U Ori with meaningful data.  RT Vir has
consistently higher values reaching $\overline{\tau}=25$.  If
$\Delta{n}$ is overestimated by a factor of 20, $\tau$ would be
reduced by $\la5$.  If $\Delta{n}$ is a
lower limit, $T_{\mathrm{x}}$ and $\tau$ are also lower limits.

\subsubsection{Maser beaming angle}
\label{saturation:beaming}
The full angular size of a cloud and the apparent angular FWHM of
maser features are approximated by $l$ and $d$ (as marked in
Fig.~\ref{F58.PS}).  The small size of $d$ means that the relative
measurement errors are large and its interpretation depends on the
internal cloud structure.  Figs~\ref{F58.PS}, \ref{94POS.PS}
and~\ref{RTSPOTPOS.PS} show that the smallest components sampling the
velocity channels near the line peak of bright features are often more
closely-spaced on the sky than the fainter components in the line
wings.  If the maser features came from completely quiescent spheres,
$d$ would be vanishingly small.  We calculated the average ratio
$d/s_0$ for each object at each epoch, where $s_0$ is the observed
size of the single maser component closest to the peak of the line
profile fitted to each feature.  This ratio is in the range 0.3 to
2.4, error-weighted mean $0.9\pm0.2$.  This is consistent with a
systematic velocity gradient within features of similar magnitude to
the thermal velocity dispersion.  

Subject to these caveats, we can obtain order-of-magnitude estimates
of the beaming angle $\Omega_{\mathrm {est}} = d^2/l^2$.  RYC99
obtained an average of $1.5\pm0.8\times10^{-3}$ sr for the 5 brightest
features in S Per, which suggested that they were on the verge of
saturation.  We obtained error-weighted mean values per epoch,
$\overline{\Omega_{\mathrm {est}}}$, in the ranges
$(0.8-6)\times10^{-3}$ sr for S Per, $(0.6-2)\times10^{-3}$ sr for U
Her and $(0.5-10)\times10^{-3}$ sr for RT Vir.  U Ori and IK Tau gave
larger values, of $(3-100)\times10^{-3}$ sr and $(10-50)\times10^{-3}$
sr, respectively.

\citet{Vlemmings05}, following \citet{Nedoluha92}, show that masers
remain unsaturated as long as $T_{\mathrm{B}}\Omega\le 10^{10}$ K sr,
for $\Gamma = 1$ s$^{-1}$.  We obtain average values of
$\overline{T_{\mathrm{B0}}}\overline{\Omega_{\mathrm {est}}} < 10^9$ K
sr for all sources apart from RT Vir, where
$\overline{T_{\mathrm{B0}}}\overline{\Omega_{\mathrm {est}}} \la
10^{11}$ K sr.  However, the rapid variability of RT Vir
(\citealt{Richards99m}; \citealt{Lekht99}) suggests that its masers
are also largely unsaturated, possible if $\Gamma>1$.
\citet{Vlemmings05a} estimate $T_{\mathrm{B}}\Omega \approx
10^{10}-10^{11}$ for S Per from line profile analysis and deduce that
the masers are mostly unsaturated, although their VLBA observations
only detected the brighter, more compact emission.

\subsubsection{Maser line narrowing}
\label{saturation:line}

 The maser FWHM linewidth is given by $\Delta{V_{1/2}} =
\Delta{V_{\mathrm{D}}}/\sqrt{1+\tau}$ \citep{Elitzur92} for unsaturated
amplification of ambient thermal radiation.  This predicts $0.25 \la
\Delta{V_{1/2}} \la 0.4$ for $\tau$ in the range shown in
Table~\ref{alpha} and $\Delta{V_{\mathrm{D}}}$ between 1.2--1.6 km
s$^{-1}$. The average measured values of $\Delta{V_{1/2}}$, given in
Table~\ref{obstab}, are greater than predicted, whilst narrower than
$\Delta{V_{\mathrm{D}}}$.  Approaching saturation causes the line
profile to re-broaden.  Alternatively,
the
observed linewidths may be explained by the velocity gradients within
the clouds. \citet{Vlemmings05a} show that, in the presence of a
velocity difference along the maser path of up to
$\Delta{V_{\mathrm{D}}}$, the unsaturated maser line width is up to
about double the width at constant velocity.

\subsubsection{Predominantly unsaturated masers}
\label{saturation:saturation}

A large value of $\Delta{n}$ provides stronger population inversion
which provides conditions for bright, unsaturated emission (see
\citet{Reid88}, their fig. 6.2), and a negative $T_{\mathrm{x}}$ of
small magnitude, approaching zero from below.
However, if the maser approaches saturation,
$\Delta{n}$ becomes small and $T_{\mathrm{x}}$ approaches $-\infty$ as
the maser emission rate becomes close to the pump rate.
  The small-magnitude negative
values we obtained for $T_{\mathrm x}$ suggest that
even the most intense H$_2$O masers in these objects are unlikely to
be strongly saturated.
Our estimates of beaming angle and line narrowing also rule out strongly
saturated water masers.  

We investigated the relationship between excitation temperature and
peak brightness temperature for each source by evaluating the slope
$A$ of the expression $\log{|T_{\mathrm{x}}|} \propto
A\log{T_{\mathrm{B0}}}$, for measurements more significant than than
3$\sigma$. This is an inverse relationship at every epoch.  RT Vir has
the shallowest slopes, averaging $-0.6\pm0.2$; the average slopes for
the other sources are U Her $-0.8\pm0.1$, S Per $-0.9\pm0.1$ and IK
Tau $-1.0\pm0.2$. U Ori has the least significant data, giving
$-0.6\pm0.5$. This shows that the brightest masers are associated with
the smallest-magnitude values of $T_{\mathrm{x}}$, consistent with
unsaturated emission.

The maser optical depth $\tau$ is equivalent to the product of the
unsaturated absorption coefficient $\kappa_0$ and the path length $l$.
We can deduce from the arguments of EHM92\footnote{note that EHM92 use
$l$ as the radius of a cloud whereas we use it as the diameter} that a
spherical maser begins to saturate when $\tau \ga 17$ and becomes
fully saturated when $\tau \ga 100$. The values of $\tau$ which we
obtain are well within the unsaturated or slightly saturated regimes.
All these results imply that circumstellar 22-GHz H$_2$O masers are
predominantly unsaturated.

 \begin{figure}
   \centering
   \includegraphics[angle=0, width=9.5cm]{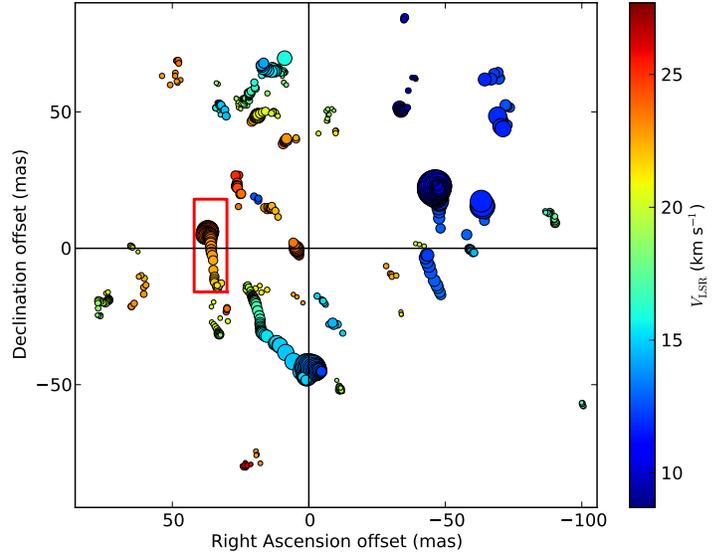}
      \caption{Each symbol represents an H$_{2}$O maser component in a
              single channel observed towards RT Vir, epoch 960524.  The
              diameter is proportional to the square root of total
              intensity. The colour is proportional to the velocity as
              shown. The boxed
              feature is shown in more detail in
              Figs.~\ref{RTSPOTPOS.PS} and~\ref{RTSELSPVELPROF.PS}.  }
         \label{RTALLPOS.PS}
   \end{figure}
\begin{figure}
 \centering
   \includegraphics[angle=0, width=9cm]{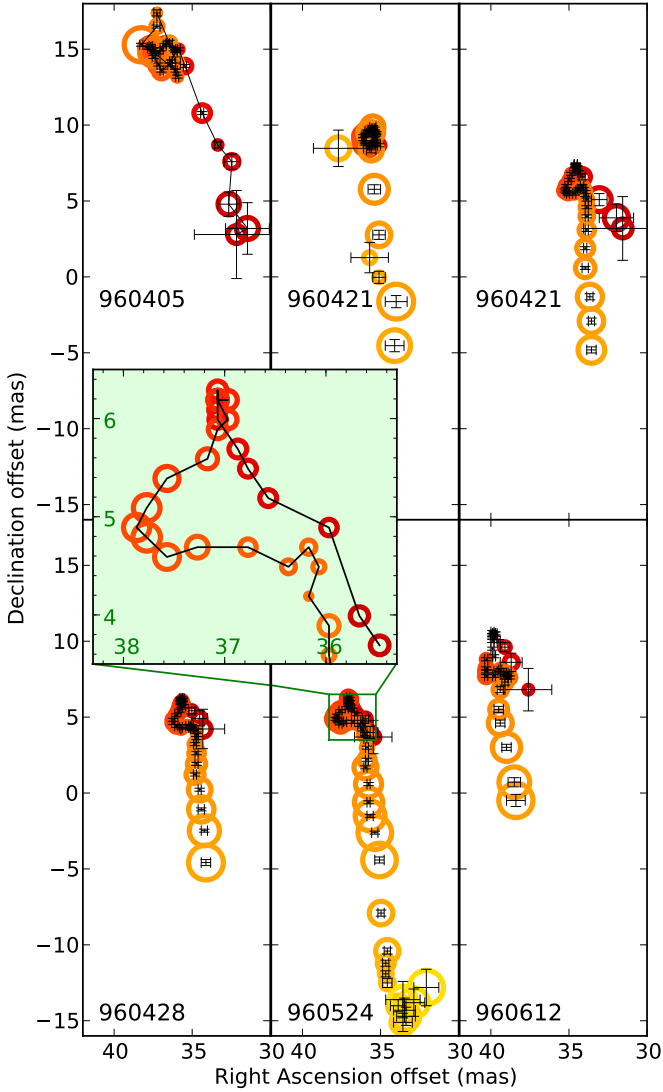}
      \caption{The angular size and distribution of the maser
 components in the RT Vir feature boxed in Fig.~\ref{RTALLPOS.PS},
 with the same colouring scheme. Each panel shows a single feature.
 The diameter of each symbol is 20\% of $s_{\sv}$ and the error bars
 represent the position offset uncertainties. The green inset shows an
 enlargement of part of the 960524 feature. }
         \label{RTSPOTPOS.PS}
   \end{figure}

 \begin{figure}
   \centering
   \includegraphics[angle=-90, width=9cm]{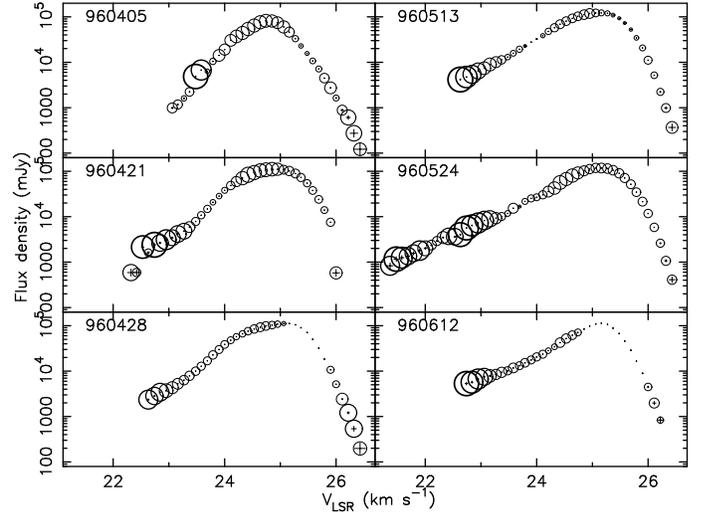}
      \caption{The line profiles of RT Vir features
  marked on Fig.~\ref{RTALLPOS.PS}.   The diameters of the
  circles and crosses are proportional to the observed angular FWHM
  $s_{\sv}$ of the maser components and to $\sigma_{s}$, respectively.
}
         \label{RTSELSPVELPROF.PS}
   \end{figure}

\section{The observed size of maser components}
\label{beaming}

The brightest components near the line centres have more compact sizes
in many of our sources, with larger components in the line wings.
Figures~\ref{ALLFEATSSPER.PS}--\ref{RTSPVELPROF.PS} show the
variations of $s_{\sv}$ across the line profiles at each epoch.
$s_{\sv}$ increases significantly from the line centres to the wings
of almost all S Per features (apart from the faintest, with larger
uncertainties). Most RT Vir features also show an inverse variation of
$s_{\sv}$ with intensity, although some components near the peaks are
anomalously swollen.  In contrast, many U Ori and U Her features have
no apparent correlation between $s_{\sv}$ and intensity.  IK Tau shows
a mixture of behaviours.
Figs.~\ref{F58.PS}, \ref{MULTIFEATSS.PS}, \ref{UOSELSPVELPROF.PS} and
\ref{RTSELSPVELPROF.PS} show the line profiles of the selected
features labelled in Figs~\ref{SPER_9499_BOXES.PS},
\ref{UORIALL000_FF.PS} and ~\ref{RTALLPOS.PS}.  These are examples of
differences in appearance which are characteristic of the three
sources.  U Her is similar to U Ori and IK Tau is somewhat similar to
RT Vir.

  Figs.~\ref{94POS.PS},
\ref{UORISPOTPOS.PS} and~\ref{RTSPOTPOS.PS} show $s_{\sv}$ as a
function of position for the selected features (at 20\% of the actual
measured size of $s_{\sv}$, in order to avoid excessive overlaps).
The error bars represent the position errors.  These plots show that
$s_{\sv}/l$ ranges from $\ll1\%$ for the peak S Per and RT Vir
components, to $>50\%$ for the components close to their line edges,
whilst  U Ori shows a less ordered scatter.

In the following sections, we compare the observed relationships
between the sizes of individual maser components and their
intensities, with the models developed by EHM92. We tentatively
investigate the effects of the sizes and velocity gradients of the
parent clouds.  A fully quantitative analysis is difficult due to the
difficulties in measuring the three-dimensional structure of the
clouds nor of the beaming angle. Nonetheless, we are able to
distinguish between the predictions for amplification- and
matter-bounded masers and identify which aspects of maser behaviour
are well-described by current theories and which confound them.

\subsection{Spherical, uniform clouds}
\label{sphere}
We start by comparing our results with the simplest model, for
spherical, density bounded clouds. EHM92 show that these would produce
amplification-bounded masers wherein the beaming angle should increase
with frequency offset from the line centre.  Initially, we assume that
all the clouds are the same size and all are spherical (i.e., that $l$
is constant within each source).

\subsubsection{Component size variation with intensity: the model}
\label{sphere:model}
The observed size $s_{\sv}$ represents the angular FWHM of the
distribution of the intensity of maser emission from water molecules
in a particular cloud in the velocity interval sampled by a single
channel.  Using expressions from \citet{Elitzur92} for the effect of
the maser amplification process in spherical clouds, the beaming solid
angle is given by
$\Omega = \pi/0.5l\kappa_{0v}
$, 
for unsaturated emission from a sphere of diameter $l$, where the unsaturated
absorption coefficient for that velocity channel is $\kappa_{0v}$. (If the
emission is significantly saturated the beaming angle is reduced by a factor of
$r_s/0.5l$ where $r_s$ is the radius of the unsaturated core.)  The observed
area is given by $A = (\pi/4\ln{2})s_{\sv}^2 = \Omega{(0.5l)^2}$, so the
apparent (unsaturated) component size should be given by
\begin{equation}
\label{s0_1}
s_{\sv}^2 \propto 0.5l/\kappa_{0v}.
\end{equation}

The intensity of unsaturated emission from  a single velocity channel though a cloud
of depth $l$ is given by
$I \approx S_0 e^{\kappa_{0v}l}$, if $I\gg S_0$, the input radiation
 (generally the ambient local thermal radiation at 22 GHz). This is
 clearly the case in CSEs where the background radiation is
 undetectable by interferometric observations on the scale of the
 masers. The stellar continuum could be amplified by a cloud along the
 line of sight, but there is no evidence for exceptionally bright
 blue-shifted peaks in our data. The volume filling factor is
 $\la1.5\%$ so cloud overlap is not likely to be significant. The
 maser component size is related to the emission intensity by
 ${I}/{S_0} \propto e^{0.5l^2/s_{\sv}^2}$ so (assuming $\ln{I} \gg
 \ln{S_0}$)
\begin{equation}
\label{I_1}
\log(s) \propto \log(\!\sqrt{0.5}l) +\alpha \log(\ln I)
\end{equation}

\subsubsection{Relationships between observed component size and
  intensity}
\label{sphere:obs}

\begin{table*}
\begin{tabular}{llcccccrrc}
\hline
Star &
Date&$\overline{T_{\mathrm{B}0}}$&$T_{\mathrm{B}0}$ min&$T_{\mathrm{B}0}$ max& $\overline{T_{\mathrm{x}}}$ & $\overline{\tau}$ &\multicolumn{1}{c}{$\alpha$}&\multicolumn{1}{c}{N$\beta$}&$\beta$\\\\
\hline
S Per  &940324&4.5$\times10^{11}$& $2.6\times10^{06}$&$3.5\times10^{13}$&--0.2  & 9$\pm$6       &$-1.6\pm0.1$& 539\,\,\,&$1.8\pm0.1$\\
 S Per &990110&2.5$\times10^{11}$& $3.6\times10^{06}$&$2.2\times10^{13}$&--0.2  &13$\pm$7       &$-1.3\pm0.1$& \emph{195}\,\,\,&\emph{1.0$\pm$ 0.1}\\
\\
U Ori  &940417&6.7$\times10^{09}$& $3.1\times10^{06}$&$9.3\times10^{10}$&--0.4  &\emph{10$\pm$3}&$ 0.6\pm0.1$& \emph{15}\,\,\,&\emph{0.8$\pm$ 0.2}\\
U Ori  &990109&2.2$\times10^{12}$& $2.7\times10^{06}$&$2.7\times10^{13}$&--0.2  &\emph{2$\pm$5} &$-3.1\pm0.2$& \emph{14}\,\,\,&\emph{0.4$\pm$ 0.1}\\
U Ori  &000410&7.3$\times10^{08}$& $2.4\times10^{06}$&$2.3\times10^{10}$&--1.7  & 8$\pm$3       &$ 1.1\pm0.2$& \emph{6}\,\,\,&\emph{0.8$\pm$ 0.3}\\
U Ori  &010506&8.2$\times10^{08}$& $3.0\times10^{06}$&$1.9\times10^{10}$&--3.2  &10$\pm$4       &$ 1.5\pm0.1$&  52\,\,\,&$0.9\pm0.2$\\
\\
U Her  &940413&3.7$\times10^{10}$& $1.3\times10^{06}$&$5.7\times10^{11}$&--0.5  &16$\pm$7       &$ 1.6\pm0.1$&  38\,\,\,&$1.0\pm0.1$\\
U Her  &000519&1.6$\times10^{11}$& $5.3\times10^{06}$&$5.3\times10^{12}$&--0.5  &13$\pm$5       &$ 2.0\pm0.1$& 101\,\,\,&$1.2\pm0.1$\\
U Her  &010427&2.4$\times10^{10}$& $2.6\times10^{06}$&$7.1\times10^{11}$&--0.4  &10$\pm$5       &$ 0.8\pm0.1$& \emph{36}\,\,\,&\emph{1.0$\pm$ 0.3}\\
\\
IK Tau &940415&9.8$\times10^{07}$& $2.2\times10^{06}$&$8.8\times10^{09}$&--4.4  &12$\pm$3       &$ 1.2\pm0.1$&\emph{141}\,\,\,&\emph{1.0$\pm$ 0.1}\\
IK Tau &000520&3.4$\times10^{09}$& $6.3\times10^{06}$&$1.4\times10^{11}$&--1.6  &14$\pm$3       &$-3.4\pm0.1$& 181\,\,\,&$0.8\pm0.1$\\
IK Tau &010427&1.0$\times10^{10}$& $6.0\times10^{06}$&$3.7\times10^{11}$&--1.2  &14$\pm$4       &$-2.5\pm0.1$&  67\,\,\,&$1.0\pm0.2$\\
\\
RT Vir &940816&5.1$\times10^{09}$& $6.2\times10^{06}$&$2.0\times10^{11}$&--1.4  &17$\pm$2       &$-1.1\pm0.1$& 199\,\,\,&$0.9\pm0.1$\\
RT Vir &960405&3.3$\times10^{09}$& $8.1\times10^{06}$&$1.2\times10^{11}$&--0.7  &21$\pm$5       &$-0.5\pm0.1$& \emph{106}\,\,\,&\emph{0.8$\pm$ 0.1}\\
RT Vir &960421&3.5$\times10^{10}$& $1.4\times10^{07}$&$8.1\times10^{11}$&--0.3  &19$\pm$4       &$ 0.1\pm0.1$& 116\,\,\,&$0.6\pm0.1$\\
RT Vir &960429&1.1$\times10^{12}$& $1.1\times10^{07}$&$5.3\times10^{13}$&--0.2  &23$\pm$5       &$-0.8\pm0.1$& 172\,\,\,&$0.8\pm0.1$\\
RT Vir &960515&8.1$\times10^{10}$& $8.7\times10^{06}$&$2.1\times10^{12}$&--0.5  &25$\pm$4       &$-0.4\pm0.1$& 142\,\,\,&$1.0\pm0.1$\\
RT Vir &960524&2.5$\times10^{10}$& $9.4\times10^{06}$&$4.4\times10^{11}$&--0.6  &25$\pm$3       &$-0.5\pm0.1$& 154\,\,\,&$0.9\pm0.1$\\
RT Vir &960612&2.2$\times10^{12}$& $7.2\times10^{07}$&$5.4\times10^{13}$&--0.1  &24$\pm$6       &$-1.0\pm0.1$& 100\,\,\,&$0.9\pm0.1$\\
\hline
\end{tabular}
\caption{Properties of H$_{2}$O masers. All temperatures are in K. The
  brightness temperature $T_{\mathrm{B}0}$ was derived using
  Equation~\ref{Tb}. $\overline{T_{\mathrm{B}0}}$ is the mean value
  for the peaks of all features in each source, given along with the
  minimum and maximum values for the feature peaks. The mean values of
  the maser excitation temperature $\overline{T_{\mathrm{x}}}$ and
  optical depth $\overline{\tau}$ were derived as described in
  Section~\ref{saturation}, which explains the uncertainties and
  dispersions; the given values are probably lower limits.  $\alpha$
  parametrises the relationship between component size and intensity
  for spherical, same-size clouds, as defined in
  Equation~\ref{I_1}. N$\beta$ is the number of components for which
  the line profile fitting was better than 3$\sigma$,
  $s_{\sv}>3\sigma_s$, $s_{\sv} > s_0$ and the component lies within a
  feature velocity span of $\le\Delta{V_{\mathrm{D}}}$.  $\beta$
  parametrises the relationship between component size, position in
  the line profile and the effect of the internal velocity gradient
  see Equation~\ref{sdlv}. Numbers in italics indicate unreliable
  results where $<30$\% of the data for that epoch meet these
  requirements.}
\label{alpha}
\end{table*}

\begin{figure}
   \centering
   \includegraphics[angle=-90, width=9cm]{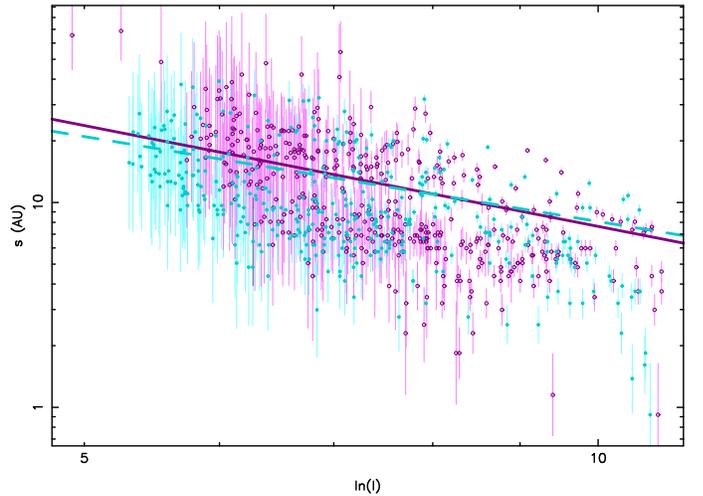}
      \caption{S Per maser component FWHM $s_{\sv}$ as a function of the natural
              logarithm of the intensity $I$ coded per epoch
              as in Fig.~\ref{UOSPOTS_I_94990001_SE3.PS}.}
         \label{SPER_SPOTS_I_9499.PS}
   \end{figure}

\begin{figure}
   \centering
   \includegraphics[angle=-90, width=9cm]{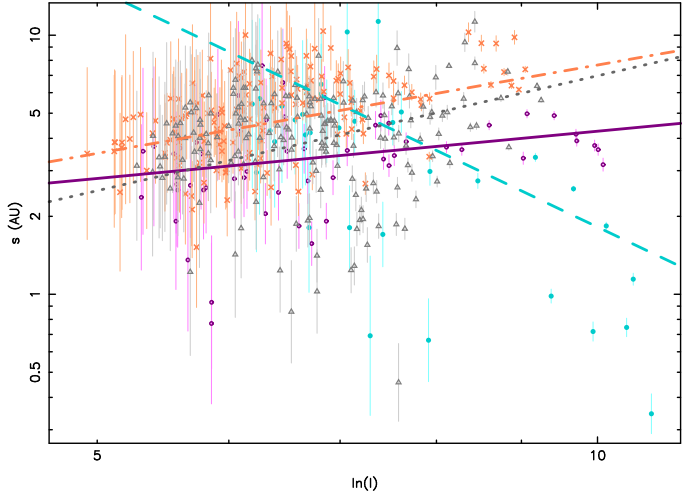}
      \caption{U Ori maser component FWHM $s_{\sv}$ as a function of the natural
              logarithm of the intensity $I$, for components where
              $s_{\sv}>3\sigma_{s}$.  The magenta (hollow), cyan
              (solid), orange (cross) and grey (triangle) symbols
              represent 1994, 1999, 2000 and 2001 results,
              respectively. The corresponding error-weighted
              least-squares fits (to all data) are shown as solid,
              dashed, dot-dashed and dotted lines. }
         \label{UOSPOTS_I_94990001_SE3.PS}
   \end{figure}

\begin{figure}
   \centering
   \includegraphics[angle=-90, width=9cm]{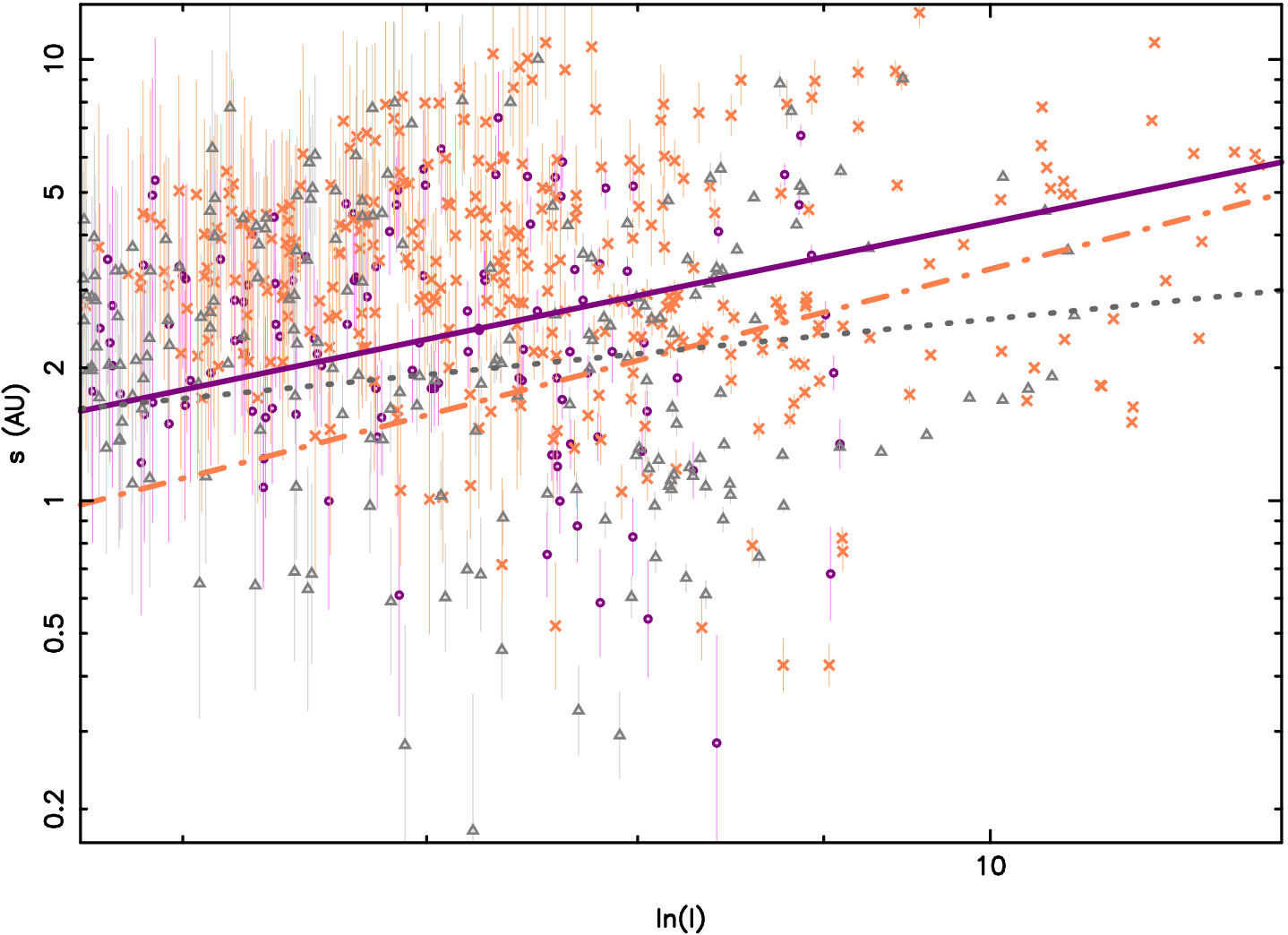} 
      \caption{U Her maser component FWHM $s_{\sv}$ as a function of the
              natural logarithm of the intensity $I$, coded per epoch
              as in Fig.~\ref{UOSPOTS_I_94990001_SE3.PS}.}
         \label{UHSPOTS_I_940001_SE3.PS}
   \end{figure}

\begin{figure}
   \centering
   \includegraphics[angle=-90, width=9cm]{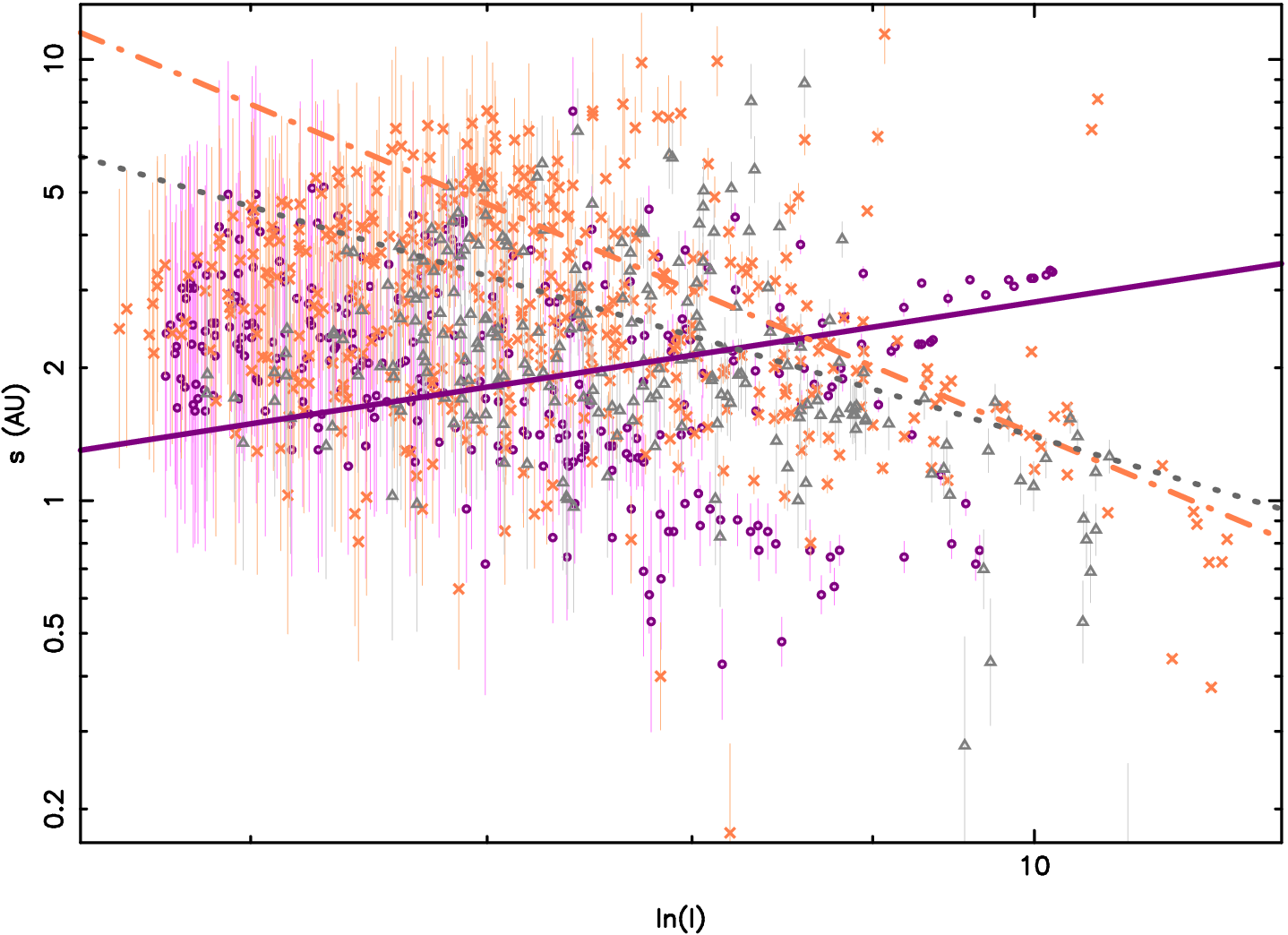}
      \caption{IK Tau maser component FWHM $s_{\sv}$ as a function of the
              natural logarithm of the intensity $I$, coded per epoch
              as in Fig.~\ref{UOSPOTS_I_94990001_SE3.PS}.}
         \label{IKSPOTS_I_940001_SE3.PS}
   \end{figure}

\begin{figure}
   \centering
   \includegraphics[angle=-90, width=9cm]{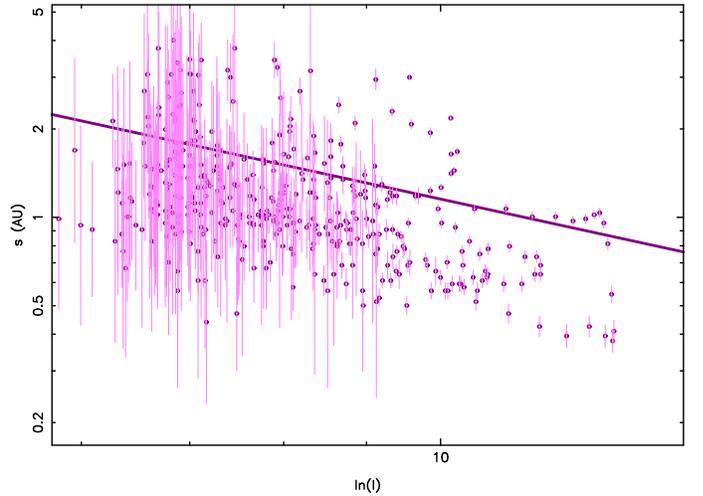}
      \caption{RT Vir maser component FWHM $s_{\sv}$ as a function of the
              natural logarithm of the intensity $I$, observed in 1994.}
         \label{RTSPOTS_I_94_SE3.PS}
   \end{figure}

\begin{figure}
   \centering
   \includegraphics[angle=-90, width=9cm]{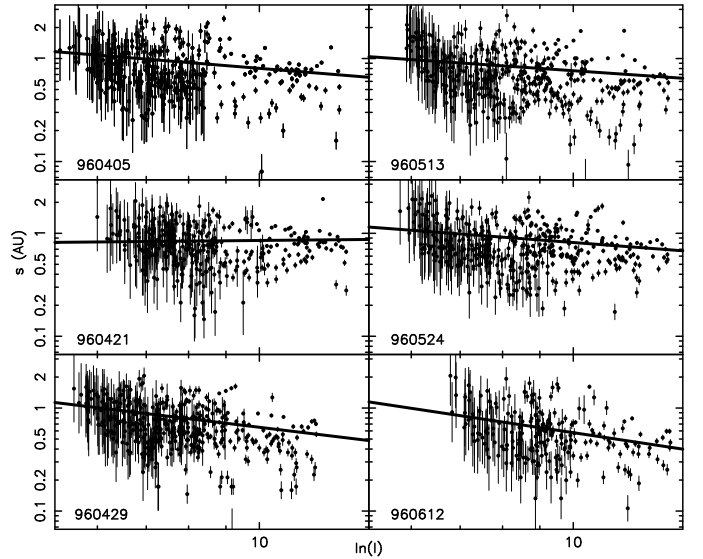}
      \caption{RT Vir maser component FWHM $s_{\sv}$ as a function of the
              natural logarithm of the intensity $I$, observed in 1996.}
         \label{RTSPOTS_I_96-6_SE3.PS}
   \end{figure}

We test the simple, spherical model for amplification-bounded masers
 (from clouds of constant size) by plotting $\log{s}$ against
 $\log(\ln{I})$. Equation~\ref{I_1} predicts that this should give a
 slope of $\alpha =
 -0.5$. Figs.~\ref{SPER_SPOTS_I_9499.PS}~--~\ref{RTSPOTS_I_96-6_SE3.PS}
 show least-squares error-weighted fits to the data for all sources.
 The values of $\alpha$ are given in Table~\ref{alpha}.  $\alpha$ is
 negative, showing that the apparent component size shrinks with
 increasing intensity, at both epochs for S Per, at 6/7 epochs for RT
 Vir, at 2/3 epochs for IK Tau and at 1/4 epochs for U
 Ori. 

Figs.~\ref{RTSPOTS_I_94_SE3.PS} and~\ref{RTSPOTS_I_96-6_SE3.PS}
 show that RT Vir most closely exhibits the behaviour predicted for
 spherical clouds, with a mean value of $\alpha$ of --0.5 for all 7
 epochs.  The behaviour is fairly consistent between epochs.  

S Per shows $\alpha<-0.5$ for both epochs, as plotted in
  Fig.~\ref{SPER_SPOTS_I_9499.PS}.  The uncertainty in $s_{\sv}$ is
  inversely proportional to $I$ and hence $\sigma_s$ is greater for
  fainter components, biasing the fit towards the contribution of the
  brighter components.  These are most likely to be approaching
  saturation, causing a steeper shrinkage of $\Omega$ with increasing
  maser amplification.  There are sufficient data for S Per to examine
  the properties for subsets.  Selecting the fainter components only,
  with $I < 0.6$ Jy, gives $\alpha = -0.6$ and $-0.4$ for the 1994 and
  1999 data respectively, close to the predicted values for
  unsaturated emission. This suggests that the steeper slopes measured
  for the whole data sets are due to saturation of the brighter
  components. This is consistent with the conclusion of RYC99 that the
  brightest features are approaching saturation.

 IK Tau also produces $\alpha<-0.5$ in 2000 and 2001, but inspection
 of Fig.~\ref{IKSPOTS_I_940001_SE3.PS} suggests a similar explanation
 as for S Per, although attempting to fit to sub-sets of the data
 produces a very large scatter.  The apparent slope of $\alpha>0$ in
 1994 is due to a small group of bright components showing anomalous
 behaviour; if these are excluded by selecting $I<7.5$ Jy, a slope of
 --1.6 is obtained, whilst for $I<0.75$ Jy, $\alpha=-0.6$, but with
 low significance since there are few points with accurate
 measurements.

U Ori and U Her have $\alpha>0$ for all but one epoch and
  Figs~\ref{UOSPOTS_I_94990001_SE3.PS}
  and~\ref{UHSPOTS_I_940001_SE3.PS} show very large scatters. This
  shows that most of their maser regions do not behave as spherical
  clouds, discussed further in Section~\ref{sec:matter}.


\subsection{Cloud size variations and velocity gradients}
\label{sec:lineprofiles}

The observed size of clouds within each CSE varies by a factor of two or more
(Table~\ref{obstab}). We  investigated whether this can be
included meaningfully in our models and attempted to allow for
the effects of velocity gradients within clouds. This is described in
Appendix~\ref{lineprofiles}. In brief,
we modified Equation~\ref{I_1} to include the change in path length
through a cloud with velocity offset. Fitting this to the data gave
slopes roughly consistent with the expected relationship between
component size and position in the line profile,
for those sources which showed the behaviour
expected from spherical clouds (i.e. $\alpha<0$ in
Equation~\ref{I_1}).  However there were very large scatters,
unsurprising because we are measuring velocity along the line of sight
but angular separation perpendicular to this.  In reality the clouds
are not perfectly spherically symmetric and we have insufficient
constraints on the three dimensional structure of individual clouds to
improve these results.  Nonetheless, insofar as the results in
Appendix~\ref{lineprofiles} are meaningful, they are consistent with
the results in Section~\ref{sphere:obs}.

\subsection{Confirmation and limitations of the predictions for
  amplification-bounded masing}

To sum up these results, the measured size of maser components has an
inverse relationship with intensity (Section~\ref{sphere}) for S Per,
RT Vir and most epochs of IK Tau, as predicted for
amplification-bounded masers. In some of these cases, the component
size shrinks more steeply than expected.  This is partly because
results are biased towards the brightest components, which may be
approaching saturation.  The remaining sources, U Ori and U Her, show
a mixture of behaviours and may be matter-bounded, see
Section~\ref{discussion}.  This first, simple investigation, of the
relationship between component size and intensity, assumed that all
the features in a given source/epoch arise from identical, quiescent
spheres.

We next attempted to investigate the behaviour of components within
features, with respect to the component at the peak of each line, for
masers meeting the amplification-bounded criterion of $s\sv > s_0$
(Appendix~\ref{lineprofiles}).  We derived a semi-empirical
relationship which included the effects of velocity gradients.  The
average ratio of the size of a component with respect to the size of
the line peak component does increase with displacement from the line
centre, but more weakly than predicted.

  EHM92 note that the beaming angle has a linear (rather than
quadratic) dependence on the ratio between the observed size and the
depth of an emission region, when these are determined by orthogonal
velocity gradients.  However, further work is needed to explore
application of this model quantitatively to our data due to the
probable unsaturated state and irregular velocity gradients. In
future, we hope to use proper motion studies which could establish the
relationship between the velocity gradients parallel and perpendicular
to the line of sight.  \emph{e}-MERLIN, will have at least double the
sensitivity of MERLIN in the same width velocity channel, and allow
sampling of any velocity span at any desirable resolution.

\section{Beaming and the properties of masing clouds}
\label{discussion}

We have compared maser properties with the predictions of maser
beaming models. We now examine the implications for the geometry of
the individual masing features and the conditions in the CSEs as a
whole.  Table~\ref{obstab} shows that $\epsilon\ge0.5$, and
$r_{\mathrm{o}}\sim(3-4)r_{\mathrm{i}}$.  Applying the expressions
given in Section~\ref{clouds:denser} suggests that if clouds are initially
spherical, expanding radially away from the central star under
radiation pressure alone, their tangential:radial aspect ratio will
remain $<2$.  Some of our results are consistent with this scenario
(Section~\ref{amp}) but in other sources some of the clouds may be
more flattened (Section~\ref{sec:matter}).

\subsection{Amplification-bounded masers}
\label{amp}
 The measured size $s_{\sv}$ of H$_{2}$O maser components decreases
with increasing maser intensity, for emission from S Per, and, for the
majority of epochs, from RT Vir and IK Tau (Section~\ref{beaming}).
The relationships roughly follow the expected relationship for maser
beaming from unsaturated, spherical masers (Equation~\ref{I_1}).
EHM92 call such masers ``amplification bounded'' since the
amplification process controls the observed size. 

The beaming angle should increase with the frequency offset from line
centre within an amplification-bounded maser feature.  The relative
observed component size, $\ln(s_{\sv}/s_0)$, is expected to be
proportional to the velocity displacement from the line peak; we also
attempt to allow for the observed velocity gradients using a
semi-empirical relationship (Equation~\ref{sdlv}).  Qualitative
agreements are obtained for sources/epochs where amplification-bounded
masing dominates but the slopes are shallower than predicted, probably
due to the over-simplified nature of the model and the non-linear
velocity gradients.

One  aspect of the behaviour of RT Vir is anomalous. Figs.~\ref{RTSPVELPROF.PS}
and~\ref{RTSELSPVELPROF.PS} show that, in some of the brightest clouds, the
components in the line wings appear relatively large, shrinking at first as
expected towards the centre of the line, but then the observed sizes re-broaden
near the peaks of the spectral features. This behaviour is very variable, e.g.
the smallest components of the bright feature at 11--13 km s$^{-1}$ appear on
the near side of the peak with respect to $V_{\star}$
(Fig.~\ref{RTSPVELPROF.PS}) at epoch 1996a, but are located on the far side at
epochs just a few weeks later.  The peak velocity with respect to $V_{\star}$
increases by $\ga1$ km s$^{-1}$ over this period.
IK Tau also shows hints of such behaviour in
Fig.~\ref{IKSPVELPROF.PS}. This spatial re-broadening cannot be
explained by current maser theory.

\subsection{Matter-bounded masers}
\label{sec:matter}
 The other two AGB stars, U Her and U Ori, and some IK Tau features,
show a very loose relationship between $s_{\sv}$ and $I$. Inspection
of their velocity profiles (Figs.~\ref{UOSPVELPROF.PS}
and~\ref{UHSPVELPROF.PS}) shows that the components in some bright
features at some epochs have the inverse relationship between observed
size and intensity expected for spherical clouds.  The components in
other features, however, show very little variation with intensity,
and some brighter features exhibit larger components than do fainter
features at the same epoch. EHM92 explain how , in `matter-bounded'
beaming, the measured maser size is expected to be independent of
frequency, i.e. location in the line profile.  This is also
consistent with the larger values of $\Omega_{\mathrm{est}}$ obtained
for U Ori than for any other source, although these have large
uncertainties (Section~\ref{saturation:beaming}).

Matter-bounded masers emanate from regions where the amplification
path along the line of sight is longer than one or both physical
dimensions of the emitting region in the plane of the sky. EHM92
develop a model for maser amplification through a flattened cloud
perpendicular to the shock direction using disc-like masers
(resembling a coin viewed edge-on), and filamentary masers, which can
be envisaged as narrow tubes of amplification pointing at the
observer.  They explain how the apparent size of such masers remains
similar to the true smaller dimension of the emission region in the
plane of the sky, with little dependence on intensity or position in
the line profile.

Figure~\ref{SPOTHIST.PS} shows that U Her and U Ori have a higher proportion of
larger components than the other AGB stars, with the peaks of the distribution
of $s_{\sv}$ comparable to $l$ at some epochs.  EHM92 predicted that
filamentary masers would exhibit an irregular distribution of components whilst
a more linear distribution is expected for disc masers. The distribution of
maser components in some features around U Ori (Figs.~\ref{UORIALL000_FF.PS}
and~\ref{UORISPOTPOS.PS}) and U Her is indeed scattered, although some of the
features which show an increase of component size towards the line peak do have
apparently linear structures, possibly due to velocity gradients.

From EHM92 equation 2.2, a masing region can behave as a filamentary
maser if its depth along the line of sight ($2l$ in the notation of
EHM92) exceeds $\sqrt{\tau}$ times its width in the perpendicular
direction ($2h$ in the notation of EHM92).  U Her and U Ori have
$\tau\la16$ (Table~\ref{alpha}), derived by assuming spherical clouds.
$\tau$ has an inverse logarithmic dependence on $|T_x|$, which in turn
has a logarithmic dependence on cloud depth (as explained in
Section~\ref{saturation:tau}).  Hence, if the depths of some clouds along
the line of sight is greater than the measured size $l$, $\tau$ will
be slightly reduced.

Taking this into account. the filamentary condition requires maser
clouds with a depth $\sim5$ times their width.  If we take
$\overline{l}$ as the width, this implies cloud depths of 12--25 AU
for U Her and U Ori, which is only half the maser shell depth
($r_{\mathrm{o}}-r_{\mathrm{i}}$).  We do
see some very long component series, e.g. the feature running from
(--20, --20) to (--10, --60) in Fig.~\ref{UORIALL000_FF.PS} epoch
1994, about 10 AU in extent.  This may be due to the effect of a
velocity gradient which allows us to see many components across the
narrow edge of a disc maser.

If clouds are initially spherical, but develop an aspect ratio
$\approx5$, this cannot be achieved by radiation pressure alone, as
explained at the start of this Section.  The other possible cause is
shock compression and U Her and U Ori are in fact the most likely
candidates for strong shocks reaching the H$_{2}$O maser region.

 The H$_2$O maser shells around U Ori and U Her have volume filling
factors 2--3 times smaller than around RT Vir and IK Tau, despite
having a higher mass loss rate than RT Vir, which implies that masing
clouds are more disrupted around U Ori and U Her.  \citet{GCVS08} and
data from the AAVSO (American association of variable star observers)
show that U Ori and U Her have regular, deep stellar pulsations with
amplitudes $\approx5$ mag, but IK Tau is a somewhat less regular
pulsator, with erratic amplitudes.  S Per and RT Vir are semi-regular
variables, with pulsation amplitudes of $<3$ and $<1$ mag,
respectively. Stronger shocks are likely to propagate out from the
stellar surfaces of U Ori and U Her, which could flatten maser clouds.
\citet{Rudnitskij00} describes signs of the impact of stellar
pulsations apparent from long-term spectral monitoring of H$_{2}$O
masers around U Ori. Moreover, both U Ori and U Her have undergone OH
maser flares likely to have originated in the inner CSE
(\citealt{Pataki74}; \citealt{Chapman85}; \citealt{Etoka97}).

%

\section{Conclusions and summary}
\label{conclusions}

We measured the properties of H$_2$O masers around S Per, U Ori, U
Her, IK Tau and RT Vir at multiple epochs using MERLIN.  
The emission in each 0.1 or 0.2 km s$^{-1}$ velocity channel is
resolved into milli-arcsec scale components, which form series in
sucessive channels generally corresponding to a single spectral
feature with a Gaussian-like profile.  Many of these features have
systematic position-velocity gradients which allow us to estimate the
physical extent of maser clouds. The number densities required for
masing and other properties show that the clouds are probably discrete
physical entities. The
average feature peak brightness temperatures are $10^8 \la
\overline{T_{\mathrm{b}}} \la 10^{12}$ K.

Most of the maser features have negative excitation temperatures close
to zero and modest optical depths, consistent with mainly unsaturated
masers.  The excess population of the upper level of the maser
transition is $<1\%$ of the combined number density of both states,
except for RT Vir, where it reaches $\sim10$\% at some epochs.  We use
this to infer maser excitation temperatures of $ -0.1 \ga
\overline{T_{\mathrm{x}}} \ga -5$ K. This leads to optical depths of
8--25 (excluding two epochs with insufficient good data).

We find an inverse relationship between the measurable size of maser
components in individual velocity channels and the intensity, for S Per
and for all but one epoch each of IK Tau and RT Vir. This is broadly
consistent with the relationship predicted by  EHM92,
$s_{\sv}\propto(\ln I)^{-0.5}$,
 for amplification-bounded masing from
uniform, spherical clouds.   The brightest
H$_2$O components around S Per and IK Tau shrink more steeply than
predicted with increasing intensity, suggesting that they are
partially saturated. In such conditions the effective observed size is
reduced proportionally to the size of the unsaturated core.  
The relationship between the relative size of each component and its
position in the line profile of amplification-bounded masers is
expected to obey $\ln(s_{\sv}/s_0)\propto 1/2
(\delta{\v}/\Delta{V_{\mathrm{D}}})^2$ (where $s_0$ is the size of the
peak component in each feature and $s_{\sv}$ is the size of the
component at velocity offset $\delta{\v}$). We find that
$\ln(s_{\sv}/s_0)$ has a weaker relationship with
$\delta{\v}$. Inspection of the plots of the variation of the RT Vir
component sizes across its line profiles shows that the brightest
components re-broaden noticeably, which cannot be explained.

Most U Ori and U Her features and some around IK Tau show contrasting
behaviour more suggestive of matter-bounded masers.  The size of maser
components is almost independent of $\ln{I}$, being generally larger
in proportion to the sizes of features ($l$).  EHM92 predicted that
observed component size is independent of position in the line profile
for matter-bounded masing.  Many features in U Her and U Ori are
likely to meet the filamentary condition, due to disc-like masers,
with an aspect ratio $\la5$.  The components of U Her and U Ori have
larger observed sizes than for the other two AGB stars, and U Ori has
the largest beaming angles.  The random spatial distribution of
components within some U Ori and U Her features is also consistent
with the predictions of EHM92 for matter-bounded masers.  These
sources have the deepest and most regular pulsation periods, so shocks
could propagate more strongly into the H$_2$O maser shell and flatten
the clouds.

The two scenarios, for amplification- and matter-bounded masers, are
compared in a cartoon in Fig.~\ref{beaming.ps}.  The upper diagrams
show how maser propagation through a sphere will give a very narrow
apparent spatial size at full width half maximum (\emph{left}) but
propagation down the long axis of a disc (\emph{right}) produces an
apparent size more similar to that of the emission region. The star
names are in order of the degree to which they present the type of
masing behaviour.  These high-resolution MERLIN observations provide
direct observational tests of the predictions of EHM92. Most
circumstellar 22-GHz H$_2$O masers behave as spherical,
amplification-bounded clouds. They are probably unsaturated, the warm
conditions and small size (relative to star-forming regions) providing
a very efficient pump which allows them to reach high maser optical
depths.  Some of the maser features round the Mira variables with the
deepest-amplitude pulsations appear to be matter-bounded, possibly
emanating from shock-flattened clouds.

\begin{acknowledgements}
We thank Indra Bains, Wouter Vlemmings and the University of
Manchester Stellar group for stimulating discussions. We also thank
Bains, Alexios Louridas and Daniel Rosa-Gonzales for the use of the
data from RYC99 and B+03.  We acknowledge with thanks the variable
star observations from the AAVSO International Database contributed by
observers worldwide and used in this research. ME acknowledges the
support of NSF grant AST-0807417. We are very grateful to the referee,
whose careful inspection of the manuscript led to great improvements
in clarity.

\end{acknowledgements}

\bibliographystyle{aa}

\bibliography{REYMaserSpotv3}

\appendix

\section{Cloud size variations and velocity gradients}
\label{lineprofiles}

The observed size of clouds within each CSE varies by a factor of two or more
(Table~\ref{obstab}). This Appendix investigates whether this can be
included meaningfully in our models, followed by an attempt to allow for
the effect of  velocity gradients. The end results are
qualitatively consistent with Section~\ref{sphere:obs} but the unknown,
probably irregular, shapes of the clouds means that we cannot obtain
precise measurements.

The effect of variations in size of spherical clouds (with purely
 thermal velocity dispersions) was tested by
 rearranging Equation~\ref{I_1} in order to plot $\log s{\sv}/l$
 against $\log(\ln I)$. We obtained slopes similar to the values of
 $\alpha$ given in Table~\ref{alpha}, or slightly steeper. All S Per,
 IK Tau and RT Vir epochs had negative slopes, averages --1.3, --1.4
 and --0.9, respectively; U Her and U Ori had large scatters with
 positive averages. These results were of lower significance than the
 evaluation of $\alpha$ described in Section~\ref{sphere:obs}. It
 seems that additional effects are more significant than variations in
 cloud size, in producing deviations from the predicted value of
 $\alpha=-0.5$. This is not surprising, since the presence of
  velocity gradients is what enables us to obtain values of
 $l$ close to the physical size of clouds.

We tried a different approach by starting with an expression
  predicting the variation of component size with intensity as a
  function of velocity offset from the peak of the line profile for
  each cloud. Initially we retain a purely thermal velocity dispersion
  as we assume that all clouds are in the same saturation state in a
  given source.  The ratio of the measured size of a maser component
  in any velocity channel, to the size of the component at the line
  peak, $s_{\sv}/s_0$, should then be a consistent function of
  velocity offset from the line centre. We take $s_0$ as the size of
  the component closest in velocity to the centre of the spectral
  profile fitted to each feature as described in Section~\ref{obs}.

  Following E92, the
  unsaturated absorption coefficient varies across the line profile of
  amplification-bounded masers as
$\kappa_{0v}/\kappa_{00} =e^{-(\delta{\v}/\Delta{V}_{\mathrm{D}})^2}
$
where the absorption coefficient is $\kappa_{00}$ at the line centre,
velocity $V_0$, and $\kappa_{0v}$ is the absorption coefficient at
velocity offset $\delta{\v}$.  The thermal line
width is a weak function of distance from the star,
$\Delta{V}_{\mathrm{D}}\propto r^{-0.2}$ (Section~\ref{clouds:profiles}). We
estimated $r$ by solving for the 3D structure of each H$_2$O maser
shell as described in \citet{Murakawa03}.

Using these relationships in Equation~\ref{s0_1}
gives an expression for $s_{\sv}$ as a function of location in the
line profile
\begin{equation}
\ln\left(\frac{s_{\sv}}{s_0}\right) \propto
\frac{1}{2}\left(\frac{\delta{\v}}{r^{-0.2}}\right)^2
\label{sdv}
\end{equation}
However, testing this relationship on our data did not show the
 expected quadratic dependence but a much weaker exponent, between 0
 and 1 if it was meaningful at all.

\begin{figure}
   \centering \includegraphics[angle=-90, width=9cm]{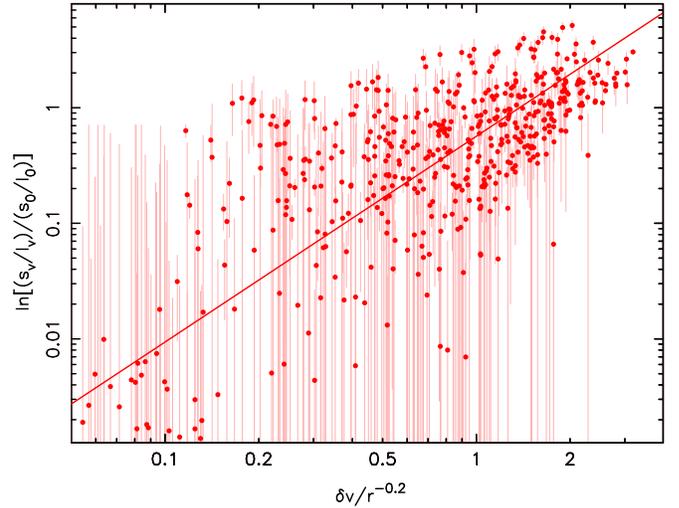}
      \caption{Variation of component size across the line profile for
    S Per in 1994.  The   size of maser components
    $s_{\sv}$ relative to the size of the component at the line centre
    $s_0$ as a function of offset from the line centre,
     weighted by the relative effective path length,
    as in Equation~\protect{\ref{sdlv}}. The  line is a least-squares
    error-weighted fit to this relationship. }
         \label{SDV.PS}
   \end{figure}

 We now attempt to extend known analytic
 relationships in order to investigate an empirical relationship
 describing the observed maser beaming by including the effects of
 velocity gradients in Equation~\ref{sdv}.

Figs.~\ref{SPER_9499_BOXES.PS} and~\ref{RTALLPOS.PS} show that many
clouds possess systematic gradients of velocity with position. The
features in Figs.~\ref{94POS.PS} A and B and~\ref{RTSPOTPOS.PS}
present orderly but non-linear structure. The progression from channel
to channel of component size and location can be followed clearly
despite twists and turns (see RT Vir inset).  Many features in IK Tau
and a few in U Her and U Ori also have systematic gradients although
the less ordered distributions seen in Figs.~\ref{UORIALL000_FF.PS}
and~\ref{UORISPOTPOS.PS} are more typical of U Ori and U Her.  The
total velocity spans ($\Delta{V}$) of features are similar in all
sources and the ratio $\Delta{V}/l$ is about 0.1--0.2 km s$^{-1}$
AU$^{-1}$ for S Per (RYC99), and an order of magnitude greater for the
smaller AGB clouds (B+03).

The angular separations between components within a cloud, especially
close to the peak, are very small and often cannot be measured
accurately.  In contrast, the velocity of each component along the
line of sight is precisely known.  We assume that the clouds are
spherical, with similar velocity-position gradients along and
perpendicular to the line of sight, allowing us to use  $l$ and $\Delta{V}$
to represent the maximum spatial and velocity extent in any
direction.  We approximate each cloud as an ensemble of overlapping
sub-spheres, each comprising molecules in a channel velocity interval
and each giving rise to a measured maser component.  We treat the
features making up the multiple-peak S Per clouds as if they were
distinct, since amplification probably only takes place along a path
length similar to $l$ (not the entire depth of the cloud, since the
velocity change is too great).
 The peak component samples the full depth, $l$, of the feature at
its centre, but the depth of a sub-sphere corresponding to a component
in the line wings is limited by the physical depth $l_{\sv}$ of a
chord through the cloud at an offset position $\delta{l}$ determined
by the local velocity gradient.  We assume that $\delta{l}$ is
proportional to the velocity offset from the feature peak,
$\delta{\v}$, for that component.
This allows us to use $\delta{l}/0.5l = \delta{\v}/0.5\Delta{V}$.
By simple trigonometry,
$(0.5l)^2 = (0.5l_{\sv})^2 + \delta{l}^2$, leading to
$l_{\sv}/l = \sqrt{1-\delta{\v}^2/(0.5\Delta{V})^2}$.

Equation~\ref{sdv} is derived from expressions for
amplification-bounded masers, so it is only applicable within features
where $s_{\sv}>s_0$, excluding much of the U Her and U Ori data from
this analysis.  We only consider features with Gaussian line profiles
fitted with better than $3\sigma$ accuracy ($N_{\mathrm{fit}}$ in
Table~\ref{obstab}), and where $\delta{\v}/0.5\Delta{V} > 1$ for both
wings. The number of components meeting these criteria, $N\beta$, is
given in Table~\ref{alpha}.

We now take $l_{\sv}$ in Equation~\ref{I_1} to represent the path
length producing the amplification of a component at velocity offset
$\delta{\v}$ of size $s_{\sv}$ and thus replace $s_{\sv}/s_0$ with
$(s_{\sv}/0.5l_{\sv} l)/(s_0/0.5l l)$ in Equation~\ref{sdv}. Using the
more precise velocity measurements as surrogates for position
measurements leads to
\begin{eqnarray}
\ln\left(\frac{s_{\sv}/l_{\sv}}{s_0/l}\right) &\propto&
\frac{1}{2}\left(\frac{\delta{\v}}{r^{-0.2}}\right)^2
\hspace*{1.4cm}
\nonumber\\ {\rm so} \hspace*{0.4cm}
\ln\left[\left(\frac{s_{\sv}}{s_0}\right)\Bigg/\sqrt{1 - \frac{\delta{\v}^2}{(0.5\Delta{V})^2}}\right]& \propto&
\frac{1}{2}\left(\frac{\delta{\v}}{r^{-0.2}}\right)^{\beta}
\label{sdlv}
\end{eqnarray}
where $\beta=2$.  

Table~\ref{alpha} gives values of $\beta$ obtained
by error-weighted least-squares fits to this relationship for all
epochs, 
  for features where $\Delta{V_{1/2}} >
3\sigma_{\Delta{V_{1/2}}}$.. Results in italics are of low significance and suggest that
this model is probably not applicable to the majority of features for
these epochs.

All the values of $\beta$ are roughly 50\% greater (i.e. closer to the
 predicted quadratic relationship) than the equivalent values of the
 exponent obtained by fitting to Equation~\ref{sdv}.

The 1994 S Per data, shown in Fig.~\ref{SDV.PS}, give
$\beta\approx1.8$, closest to the expected value.  The other sources
with significant measurements have $\beta\approx0.9$.  Many factors
could contribute to the lower values of $\beta$, such as deviations
from spherical symmetry and non-linear velocity gradients.  The closer
angular spacing seen in many cases for brighter components compared
with the displacement of the faintest from the line centre, means that
$\delta{\v}$ is not a perfect surrogate for $\delta{l}$.

\end{document}